\documentclass[12pt]{article}
\begin{document}
\title{A concept of polymeric networks with sliding junctions
for the time-dependent response of filled elastomers}

\author{Aleksey D. Drozdov and Al Dorfmann\\
Institute of Structural Engineering\\
82 Peter Jordan Street\\
A--1190 Vienna, Austria}
\date{}
\maketitle

\begin{abstract}
Constitutive equations are derived for the time-dependent
behavior of particle-reinforced elastomers at isothermal loading
with finite strains.
A rubbery polymer is modelled as a network of macromolecules
bridged by junctions, where the junctions can slip with respect to
the bulk material under straining.
A filled rubbery compound is thought of as an ensemble of meso-regions
where sliding occurs with different rates.
Stress--strain relations for a particle-reinforced rubber
are developed by using the laws of thermodynamics.
For uniaxial tensile relaxation tests, these equations are determined by
three adjustable parameters.
To find the experimental constants, three series of tests
are performed at longitudinal strains in the range from 100 to 250~\%.
By fitting observations, the effects of pre-loading and thermal
recovery are analyzed on the nonlinear viscoelastic response of
natural rubber reinforced with carbon black.
\end{abstract}
\vspace*{10 mm}

\newpage

\section{Introduction}

This paper is concerned with modelling the time-dependent behavior
of particle-reinforced elastomers at isothermal loading with finite strains.
The viscoelastic and viscoplastic responses of filled rubbers
have been a focus of attention in the past decade, which may be
explained by numerous applications of rubber-like materials in
industry (vehicle tires, shock absorbers, earthquake bearings,
seals, flexible joints, solid propellants, etc.), see, e.g.,
Govendjee and Simo (1992),
Johnson and Stacer (1993),
H\"{a}usler and Sayir (1995),
Johnson et al. (1995),
Aksel and H\"{u}bner (1996),
Holzapfel and Simo (1996),
Lion (1996, 1997, 1998),
Spathis (1997),
Bergstr\"{o}m and Boyce (1998),
Ha and Schapery (1998),
Kaliske and Rothert (1998),
Reese and Govindjee (1998),
Septanika and Ernst (1998),
Miehe and Keck (2000),
Wu and Liechti (2000),
Haupt and Sedlan (2001), to mention a few.
Although the number of publications on this subject is rather large,
some important issues remain, however, obscure.

One of the main questions to be resolved deals with the physical nature of
the time-dependent response of rubber-like materials.
It is worth noting that there is no agreement in the literature about
the micro-mechanisms for stress relaxation in unfilled elastomers.

The viscoelastic behavior of rubbery polymers is conventionally modelled
within the concept of temporary networks (Green and Tobolsky, 1946).
A polymer is treated an a network of macromolecules bridged by
junctions (chemical and physical cross-links, entanglements and inter-chain links
induced by van der Waals forces).
With reference to the affinity hypothesis, junctions are assumed to be rigidly
connected with the bulk medium, whereas the time-dependent response
is associated with breakage of active strands (whose ends are linked
to separate junctions) and reformation of dangling strands (where stresses
totally relax before these chains merge with the network).

Another scenario for stress relaxation in rubbery polymers is based
of the concept of non-affine junctions that presumes the network of
macromolecules to be permanent and explains the time-dependent response
at the macro-level by slippage of junctions with respect to the bulk
material at the micro-level.
This approach was first suggested by Phan-Thien and Tanner (1977)
and Phan-Thien (1978) for polymer melts and was applied
by Desmorat and Cantournet (2001) to study the behavior of filled elastomers.

A version of the concept of non-affine networks, where junctions do not
slide, but polymeric chains are allowed to slip with respect to entanglements
was proposed by Miehe and co-authors for the analysis of stresses
in rubbery vulcanizates (Miehe, 1995; Miehe and Keck, 2000, Lulei and Miehe, 2001).
A similar model (where chains slip with respect to filler particles
instead of entanglements) was recently introduced by Hansen (2000)
with reference to results of molecular dynamics simulations and atomic force
microscopy measurements.

The above theories concentrate on relatively slow relaxation processes
when stresses relax at the time-scale of structural rearrangement
in the entire macromolecule.
A relatively fast relaxation process (whose characteristic time is associated
with the time for rearrangement of a mechanical segment in a chain)
was studied by Drozdov (2001) by using a micro-mechanical scenario
for the time-dependent response of rubbery polymers similar to the
tube model for concentrated polymeric solutions and melts (Doi and Edwards, 1986).
The difference between the two approaches is that the conventional tube
theory is based on the reptation concept for polymeric chains (this kind of
diffusive motion is prevented by cross-links in rubbery vulcanizates),
whereas the concept of wavy tubes ascribes the viscoelastic behavior
of rubbers at the macro-level to mechanically induced changes
in the concentration of regions with high molecular mobility in an individual
strand at the micro-level.
A transition mechanism from relaxation processes at the length-scale
of a statistical segment to those at the length-scale of a chain in a polymer
melt was recently discussed by Herman (2001).

This brief discussion of basic concepts in finite viscoelasticity
and viscoplasticity of rubber-like materials demonstrates that
our knowledge of micro-mechanisms for the time-dependent response
of unfilled elastomers is far away from being exhausted.
The physics of the viscoelastic and viscoplastic behavior of
particle-reinforced rubbers remains rather unclear.
This conclusion is confirmed by a number of micro-mechanical theories
that have been proposed in the past decade to describe the viscoelastic
response of filled elastomers.

On the one hand,  Aksel and H\"{u}bner (1996),
Kaliske and Rothert (1998) and Hansen (2000) explained the
time-dependent behavior of particle-reinforced rubbers by sliding
of polymeric chains along and their de-wetting from the surfaces
of filler clusters.
This concept goes back to the model proposed by Dannenberg (1975),
which, in turn, is equivalent to the neglect of the ability
of a host matrix to relax, the hypothesis suggested about half a century ago
by Bueche (1960, 1961) to explain the Mullins effect.

On the other hand, Liang et al. (1999), Clarke et al. (2000) and
Leblanc and Cartault (2001) evidenced that the effect of particles
and their aggregates on the viscoelastic response of rubbery compounds
is not crucial.
Leblanc and Cartault (2001) performed dynamic shear tests on
uncured styrene--butadiene rubber reinforced with various amounts
of carbon black and silica and reported rather limited time-dependent
responses of the compounds.
These observations allowed them to infer that inter-molecular interactions
play the key role in the viscoelastic
behavior, whereas the influence of filler is of second importance.
Liang et al. (1999) reported experimental data in dynamic tensile tests
on ternary composite of polypropylene and EPDM (ethylene propylene diene monomers)
elastomer filled with glass beads.
They demonstrated that treatment of beads by silane coupling agent did not
affect mechanical damping and that the loss modulus of the compound
with untreated beads decreased with the concentration of filler
(in contrast with the de-wetting concept).
Clarke et al. (2000) carried out tensile relaxation tests on two
elastomers (natural polyisoprene rubber and synthetic polyacrylate rubber)
and revealed significant relaxation of stresses in the unfilled natural rubber.

Our experimental data in tensile relaxation tests on natural rubber
reinforced with high abrasion furnace black demonstrated comparable
decreases in the longitudinal stress for a filled [45 phr (parts per
hundred parts of rubber) of carbon black (CB)] and an unfilled [1 phr of CB]
compounds (Drozdov and Dorfmann, 2001).
Because the amount of stress relaxed during 1 hour at room temperature
in the filled specimens exceeded that for the unfilled samples
(approximately by twice), we conclude that a micro-mechanism for
the viscoelastic response of filled elastomers is in between
the above two extreme concepts:
the relaxation process in particle-reinforced rubbers may be associated
with rearrangement of macromolecules in the matrix,
but the rearrangement process is strongly affected by local inhomogeneities
induced by the presence of filler.

In accord with this picture, we model a filled rubber as a
network of macromolecules bridged by junctions.
To account for the effect of filler particles and their agglomerates
on the time-dependent response of the network in a tractable way,
the network is treated as an ensemble of meso-domains, where rearrangement
of strands occurs with different (mutually independent) rates.
The characteristic size of these regions (roughly estimated as several
hundreds of nm) is assumed to substantially exceed the radius of gyration
for macromolecules and the average radius of particles, on the one hand,
and to be noticeably less than the dimensions of a macro-specimen.
The ensemble of meso-domains reflects local inhomogeneities
in the rubber-like material caused by impurities,
non-homogeneity in the spatial distribution of a cross-linker and a filler,
segregation of short chains to interfaces between the filler clusters
and the host matrix (Carlier et al., 2001),
local fluctuations of elastic moduli associated with the presence of hard
cores around filler particles and their clusters,
as well as the enhancement of the rearrangement process for strands
in the close vicinities of the filler aggregates (Karasek and Sumita, 1996).
The work is confined to filled rubbers with the volume fractions
of particles, $\varphi$, below the percolation threshold, which implies that
phenomena associated with aggregation of clusters into an infinite secondary
network (Yatsuyanagi et al., 2001) are not taken into account.

With reference to the concept of sliding junctions, different meso-regions
are characterized by different potential energies, $\omega$,
for slippage of junctions with respect to the bulk material.
The rate of slippage, $\Gamma$, is expressed in terms of the potential
energy, $\omega$, by the Boltzmann formula.
This implies that the sliding process is entirely determined by
the attempt rate, $\Gamma_{0}$, and the distribution function, $p(\omega)$,
of potential energies for sliding.

To account for the influence of strain intensity on the rate of
stress relaxation, we distinguish between active meso-domains,
where junctions can freely slide,
and passive meso-regions, where sliding is prevented by
surrounding macromolecules (because local concentrations
of cross-links and van der Waals links between strands
in these domains exceed their average values).
Under stretching, some van der Waals links between strands break
in passive meso-domains, which implies that these regions become active.
The ability of a passive meso-region to be activated is determined
by a measure of damage, $\alpha$, that determines the relative
number of extra links between macromolecules which do not permit
the sliding process to start.
The present study focuses on loading programs in which passive
meso-regions are transformed into the active state when the stress
intensity grows.
Healing of inter-chain links that results in the inverse transformation
at unloading is beyond the scope of this paper.

The objective of this paper is three-fold:
\begin{enumerate}
\item
To develop constitutive equations for the time-dependent response
of a particle-reinforced elastomer modelled as an ensemble of meso-regions
with various potential energies for sliding of junctions.

\item
To report experimental data in three series of tensile relaxation tests
on CB filled natural rubber at various longitudinal strains in the range from 100
to 250 \%.
The first series of tests was carried out on virgin specimens,
in the second series the samples were initially pre-loaded,
whereas the third series dealt with the same specimens after thermal
recovery at an elevated temperature.

\item
To find adjustable parameters in the stress--strain relations by fitting
observations and to rationalize the influence of pre-loading and thermal
recovery on the nonlinear viscoelastic behavior of filled elastomers
in terms of the constitutive model.
\end{enumerate}

The study is organized as follows.
Sliding of junctions in active meso-regions and transition of passive
meso-domains into the active state are discussed in Section 2.
Kinematic equations for slippage of junctions with respect to the
bulk medium are developed is Section 3.
The mechanical energy of a particle-reinforced elastomer
is determined in Section 4.
Stress--strain relations are derived in Section 5 by using the
laws of thermodynamics.
Uniaxial extension of a specimen is analyzed in Section 6.
Section 7 deals with the description of the experimental procedure.
Adjustable parameters in the constitutive equations are determined
by fitting observations in Section 8, where a physical interpretation
is provided for the findings.
Some concluding remarks are formulated in Section 9.

\section{Deformation of a filled elastomer}

A particle-reinforced elastomer is modelled as an ensemble of meso-domains
(with arbitrary shapes) where junctions can slip with respect to
the bulk material.
Sliding of junctions is assumed to occur independently in different
meso-regions.

A meso-domain is characterized by its potential energy $\omega$,
which is defined as follows. Slippage of junctions with respect to
their positions in the bulk material is thought of as a viscous flow
with the constitutive equation
\begin{equation}
\sigma=\eta \dot{e},
\end{equation}
where $\sigma$ is the stress, $\dot{e}$ is the strain rate, and $\eta>0$
is some viscosity.
The parameter $\eta$ in Eq. (1) is given by
\begin{equation}
\eta=\frac{G}{\Gamma},
\end{equation}
where $G$ stands for an elastic modulus,
and $\Gamma$ is the rate of sliding.
For definiteness, we suppose that $G$ coincides for all meso-domains,
whereas $\Gamma$ accepts different values for different regions.
The Boltzmann formula is adopted for the rate of sliding
\begin{equation}
\Gamma=\Gamma_{0}\exp \Bigl (-\frac{\bar{\omega}}{k_{B}T}\Bigr ),
\end{equation}
where $\bar{\omega}$ is the potential energy of a meso-region, $T$
is the absolute temperature, $k_{B}$ is Boltzmann's constant, and
the pre-factor $\Gamma_{0}$ weakly depends on temperature.
The attempt rate $\Gamma_{0}$ equals the rate of sliding for junctions
at an elevated temperature ($T\to \infty$).
Introducing the dimensionless potential energy,
\[
\omega=\frac{\bar{\omega}}{k_{B}T_{0}},
\]
where $T_{0}$ is some reference temperature, and disregarding the
effect of small temperature increments, $\Delta T=T-T_{0}$, on
the sliding process, we find from Eq. (3) that
\begin{equation}
\Gamma=\Gamma_{0}\exp (-\omega ).
\end{equation}
Equation (4) expresses the rate of slippage of junctions with respect
to the bulk medium, $\Gamma$, in a meso-region in terms of
its potential energy, $\omega$.

The potential energy $\omega$ characterizes the rate of sliding of
junctions in a meso-domain where the sliding process is not affected
by inter-chain interactions.
The neglect of interaction between polymeric chains is a conventional
hypothesis in the entropic theory of rubber elasticity (Treloar, 1975),
where the number of available configurations for a chain
is determined under the assumption that surrounding
macromolecules do not restrict the chain geometry.
For a filled elastomer, interactions between polymeric chains
may be disregarded only for a part of the ensemble of meso-domains.
In the other part of meso-regions these interactions prevent sliding
of junctions until micro-strains become sufficiently large to
break inter-chain links (these links reflect van der Waals' forces
between strands) that do not allow the sliding process to start.

With reference to this picture, two groups of meso-domains in an ensemble
are distinguished:
\begin{enumerate}
\item
active regions where sliding of junctions is unrestricted,

\item
passive domains where slippage of junctions is inhibited by
inter-chain interactions.
\end{enumerate}
The sliding process in an active meso-region is entirely determined
by the potential energy, $\omega$, of this domain.
To characterize the ability of a passive meso-domain to become
active, we introduce an additional variable, $\alpha$, which is thought of
as a measure for mechanically induced damage of inter-chain interactions.

Any interaction between polymeric chains that prevents slippage
of junctions with respect to the bulk material may be thought of
as a temporary link between macromolecules.
The weakening of interaction between strands driven by mechanical
factors is modelled as rupture of temporary links under loading.

Assuming the number of meso-regions with potential energy $\omega$ per unit mass
of an elastomer to be rather large,
we denote by $\langle M(\omega)\rangle$ the average number of inter-chain
links per unit domain before loading,
and by $M(t,\omega)$ the number of these links in a meso-domain
in the deformed state at time $t$ (the instant $t=0$
corresponds to the time when external forces are applied to a specimen).
The measure of damage, $\alpha$, is defined as the
ratio of these quantities,
\begin{equation}
\alpha(t,\omega)=\frac{M(t,\omega)}{\langle M(\omega)\rangle }.
\end{equation}
Unlike conventional approaches to fracture of elastomers, where the
initial measure of damage is assumed to vanish, whereas fracture
of a macro-sample is associated with some positive value of the
damage variable (conventionally, unity), Eq. (5) implies that $\alpha$
is non-negative in the stress-free state, and it decays to zero
when a passive meso-region is transformed into an active one.

It follows from Eq. (5) that the initial value of the damage variable,
$\alpha$, may be either higher or lower than 1 in passive
meso-domains, whereas the average value of $\alpha(0,\omega)$
equals unity.
The condition $\alpha(0,\omega)<1$ means that a passive meso-domain
is initially damaged.
This damage is associated with breakage of inter-chain interactions
in meso-regions during the preparation process (at the stage
of milling of a rubbery compound).
On the other hand, the inequality $\alpha(0,\omega)>1$ implies that
the number of links between macromolecules in a passive meso-domain
in a virgin specimen exceeds its average value, which may be ascribed
to local inhomogeneities in the distribution of a cross-linker.

In accord with this picture, the distribution of passive meso-domains
is described by the function $\Xi_{\rm p}(t,\alpha,\omega)$
that equals the current number of passive meso-regions (per unit mass)
with the potential energy $\omega$ where the damage variable
equals $\alpha$.
Summing the number of meso-domains with various measures of damage,
$\alpha$, we find the concentration of passive regions with potential
energy $\omega$,
\begin{equation}
X_{\rm p}(t,\omega)=\int_{0}^{\infty} \Xi_{\rm p}(t,\alpha,\omega)d\alpha.
\end{equation}
The number of passive domains per unit mass of an elastomer, $N_{\rm p}$,
equals the sum of the numbers of passive meso-regions with
various potential energies, $\omega$,
\begin{equation}
N_{\rm p}(t)=\int_{0}^{\infty} X_{\rm p}(t,\omega) d\omega.
\end{equation}
Denote by $X_{\rm a}(t,\omega)$ the current number of active meso-domains
with potential energy $\omega$.
The quantities $X_{\rm a}$ and $X_{\rm p}$ are connected by the
conservation law
\begin{equation}
X_{\rm a}(t,\omega)+X_{\rm p}(t,\omega)=X(\omega),
\end{equation}
where $X(\omega)$ is the concentration of meso-regions with potential
energy $\omega$.
Differentiation of Eq. (8) with respect to time implies that
\begin{equation}
\gamma(t,\omega)=\frac{\partial X_{\rm a}}{\partial t}(t,\omega)
=-\frac{\partial X_{\rm p}}{\partial t}(t,\omega),
\end{equation}
where $\gamma(t,\omega)$ is the rate of transformation of passive
meso-domains into active ones.
Integrating Eq. (8) over $\omega$ and using Eq. (7), we find that
\begin{equation}
N_{\rm a}(t)+N_{\rm p}(t)=N,
\end{equation}
where
\begin{equation}
N_{\rm a}(t)=\int_{0}^{\infty} X_{\rm a}(t,\omega) d\omega
\end{equation}
is the current concentration of active meso-domains and
\[ N=\int_{0}^{\infty} X(\omega) d\omega \]
is the total number of meso-regions per unit mass.

\section{Kinematic relations}

Let ${\bf r}_{0}$ be the radius vector of an arbitrary junction
in the reference state,
${\bf r}(t,{\bf r}_{0})$ its radius vector in the deformed state
at time $t$,
and ${\bf r}_{\rm u}(t,{\bf r}_{0})$ the radius vector of this point
in the unloaded state.
In what follows, the argument ${\bf r}_{0}$ of the functions
${\bf r}$ and ${\bf r}_{\rm u}$ is omitted for the sake of simplicity.

The unloaded state of a meso-region characterizes changes in
the positions of junctions at time $t$ driven of their slippage
with respect to the bulk material (i.e., by slippage with respect to
the initial configuration of a specimen).
The displacement vector for transition of a junction from its
reference state to the unloaded state is given by
\[
{\bf u}_{\rm u}(t)={\bf r}_{\rm u}(t)-{\bf r}_{0}.
\]
Denote by ${\mathbf \nabla}_{0}{\bf r}(t)$ the deformation
gradient for transition from the reference state to the actual
state,
by ${\mathbf \nabla}_{0}{\bf r}_{\rm u}(t)$ the deformation
gradient for transition from the reference state to the unloaded
state,
and by ${\mathbf \nabla}_{\rm u}(t){\bf r}(t)$ the
deformation gradient for transition from the unloaded state to the
deformed state.
Here ${\mathbf \nabla}_{0}$ and ${\mathbf \nabla}_{\rm u}(t)$
are the gradient operators in the initial and unloaded configurations,
respectively.
With reference to Drozdov (1996), the ``left" gradient operators are employed:
the image, $d{\bf r}(t)$, in the actual state of an infinitesimal vector,
$d{\bf r}_{0}$, in the reference state reads
\[
d{\bf r}(t)=d{\bf r}_{0}\cdot {\mathbf \nabla}_{0}{\bf r}(t),
\]
where the dot stands for inner product.
The deformation gradients are connected by the equation
\begin{equation}
{\mathbf \nabla}_{0}{\bf r}(t)=
{\mathbf \nabla}_{0}{\bf r}_{\rm u}(t)\cdot
{\mathbf \nabla}_{\rm u}(t){\bf r}(t),
\end{equation}
which expresses the chain rule for differentiation of the vector
function ${\bf r}={\bf r}(t,{\bf r}_{0})$ with respect to ${\bf r}_{0}$.

The right Cauchy deformation tensor, ${\bf C}(t)$, for transition
from the initial state to the deformed state is determined by the
formula
\begin{equation}
{\bf C}(t)={\mathbf \nabla}_{0}{\bf r}(t)\cdot
\Bigl [{\mathbf \nabla}_{0}{\bf r}(t)\Bigr ]^{\top},
\end{equation}
where $\top$ denotes transpose.
By analogy with Eq. (13), we introduce the right Cauchy deformation
tensor for transition from the reference state to the unloaded state,
\begin{equation}
{\bf C}_{\rm u}(t)={\mathbf \nabla}_{0}{\bf r}_{\rm u}(t)\cdot
\Bigl [{\mathbf \nabla}_{0}{\bf r}_{\rm u}(t)\Bigr ]^{\top},
\end{equation}
and the right Cauchy deformation tensor for transition from the
unloaded state to the actual state,
\begin{equation}
{\bf C}_{\rm e}(t)={\mathbf \nabla}_{\rm u}(t){\bf r}(t)\cdot
\Bigl [{\mathbf \nabla}_{\rm u}(t){\bf r}(t)\Bigr ]^{\top}.
\end{equation}
It follows from Eqs. (12), (13) and (15) that
\begin{eqnarray}
{\bf C}_{\rm e}(t) &=&
\Bigl [ {\mathbf \nabla}_{0}{\bf r}_{\rm u}(t)\Bigr ]^{-1}\cdot
{\mathbf \nabla}_{0}{\bf r}(t)\cdot
\Bigl [ {\mathbf \nabla}_{0}{\bf r}(t)\Bigr ]^{\top} \cdot
\Bigl [{\mathbf \nabla}_{0}{\bf r}_{\rm u}(t)\Bigr ]^{-\top}
\nonumber\\
&=& \Bigl [ {\mathbf \nabla}_{0}{\bf r}_{\rm u}(t)\Bigr ]^{-1}\cdot
{\bf C}(t)\cdot \Bigl [{\mathbf \nabla}_{0}{\bf r}_{\rm u}(t)
\Bigr ]^{-\top}.
\end{eqnarray}
The left Cauchy deformation tensor for transition from the
reference state to the deformed state reads
\begin{equation}
{\bf B}(t)= \Bigl [{\mathbf \nabla}_{0}{\bf r}(t)\Bigr ]^{\top}
\cdot{\mathbf \nabla}_{0}{\bf r}(t),
\end{equation}
whereas the left Cauchy deformation tensor for transition from the
unloaded state to the actual state is given by
\begin{equation}
{\bf B}_{\rm e}(t)=
\Bigl [{\mathbf \nabla}_{\rm u}(t){\bf r}(t)\Bigr ]^{\top}
\cdot{\mathbf \nabla}_{\rm u}(t){\bf r}(t).
\end{equation}
Equations (12), (14) and (18) imply that
\begin{eqnarray}
{\bf B}_{\rm e}(t) &=&
\Bigl [{\mathbf \nabla}_{0}{\bf r}(t)\Bigr ]^{\top}\cdot
\Bigl [{\mathbf \nabla}_{0}{\bf r}_{\rm u}(t)\Bigr ]^{-\top}
\cdot \Bigl [{\mathbf \nabla}_{0}{\bf r}_{\rm u}(t)\Bigr ]^{-1}
\cdot {\mathbf \nabla}_{0}{\bf r}(t)
\nonumber\\
&=&  \Bigl [{\mathbf \nabla}_{0}{\bf r}(t)\Bigr ]^{\top}\cdot
{\bf C}_{\rm u}^{-1}(t) \cdot
{\mathbf \nabla}_{0}{\bf r}(t).
\end{eqnarray}
Our purpose now is to determine the derivatives of the principal
invariants, $I_{k}$ ($k=1,2$), of the tensor ${\bf C}_{\rm e}(t)$
with respect to time.
Differentiation of Eq. (16) results in
\begin{eqnarray}
\frac{d {\bf C}_{\rm e}}{dt}(t) &=& \frac{d}{dt}
\Bigl [ {\mathbf \nabla}_{0}{\bf r}_{\rm u}(t)\Bigr ]^{-1}
\cdot {\bf C}(t)\cdot
\Bigl [{\mathbf \nabla}_{0}{\bf r}_{\rm u}(t)\Bigr ]^{-\top}
\nonumber\\
&& + \Bigl [ {\mathbf \nabla}_{0}{\bf r}_{\rm u}(t)\Bigr ]^{-1}
\cdot \frac{d {\bf C}}{dt} (t)\cdot
\Bigl [{\mathbf \nabla}_{0}{\bf r}_{\rm u}(t)\Bigr ]^{-\top}
\nonumber\\
&& + \Bigl [ {\mathbf \nabla}_{0}{\bf r}_{\rm u}(t)\Bigr ]^{-1}
\cdot {\bf C}(t)\cdot \frac{d}{dt}
\Bigl [{\mathbf \nabla}_{0}{\bf r}_{\rm u}(t)\Bigr ]^{-\top}.
\end{eqnarray}
Bearing in mind that
\[ \frac{d{\bf r}_{\rm u}}{dt}(t)={\bf v}_{\rm u}(t), \]
where ${\bf v}_{\rm u}(t)$
is the velocity vector for sliding of junctions,
and using the chain rule for differentiation, we obtain
\begin{equation}
\frac{d}{dt}\Bigl [ {\mathbf \nabla}_{0}{\bf r}_{\rm u}(t)\Bigr ]
={\mathbf \nabla}_{0}{\bf v}_{\rm u}(t)
={\mathbf \nabla}_{0}{\bf r}_{\rm u}(t)\cdot {\bf L}_{\rm u}(t),
\end{equation}
where
\[
{\bf L}_{\rm u}(t)={\mathbf \nabla}_{\rm u}(t){\bf v}_{\rm u}(t).
\]
Taking into account that
\[
\frac{d}{dt} \Bigl [ {\mathbf \nabla}_{0}{\bf r}_{\rm u}(t)\Bigr ]^{-1}
=-\Bigl [ {\mathbf \nabla}_{0}{\bf r}_{\rm u}(t)\Bigr ]^{-1}
\cdot \frac{d}{dt}\Bigl [ {\mathbf \nabla}_{0}{\bf r}_{\rm u}(t)\Bigr ]
\cdot\Bigl [ {\mathbf \nabla}_{0}{\bf r}_{\rm u}(t)\Bigr ]^{-1},
\]
we find from Eq. (21) that
\begin{equation}
\frac{d}{dt} \Bigl [ {\mathbf \nabla}_{0}{\bf r}_{\rm u}(t)\Bigr ]^{-1}
=-{\bf L}_{\rm u}(t)\cdot
\Bigl [ {\mathbf \nabla}_{0}{\bf r}_{\rm u}(t)\Bigr ]^{-1}.
\end{equation}
It follows from Eq. (13) that the derivative of the Cauchy
deformation tensor ${\bf C}(t)$ reads
\begin{equation}
\frac{d{\bf C}}{dt}(t)={\mathbf \nabla}_{0}{\bf v}(t)\cdot
\Bigl [{\mathbf \nabla}_{0}{\bf r}(t)\Bigr ]^{\top}
+ {\mathbf \nabla}_{0}{\bf r}(t)\cdot
\Bigl [{\mathbf \nabla}_{0}{\bf v}(t)\Bigr ]^{\top},
\end{equation}
where
\[ {\bf v}(t)=\frac{d{\bf r}}{dt}(t) \]
is the velocity vector.
Taking into account that
\[
{\mathbf \nabla}_{0}{\bf v}(t)={\mathbf \nabla}_{0}{\bf r}(t)\cdot
{\mathbf \nabla}(t){\bf v}(t),
\]
where ${\mathbf \nabla}(t)$ is the gradient operator in the
deformed configuration,
we arrive at the formula
\begin{equation}
\frac{d{\bf C}}{dt}(t)=2 {\mathbf \nabla}_{0}{\bf r}(t)\cdot
{\bf D}(t)\cdot [{\mathbf \nabla}_{0}{\bf r}(t)\Bigr ]^{\top},
\end{equation}
where
\[
{\bf L}(t)={\mathbf \nabla}(t){\bf v}(t)
\]
is the velocity gradient and
\[
{\bf D}(t)=\frac{1}{2}\Bigl [ {\bf L}(t)+{\bf L}^{\top}(t)\Bigr ],
\]
is the rate-of-strain tensor.
Equations (16), (20), (22) and (24) result in
\begin{eqnarray}
\frac{d{\bf C}_{\rm e}}{dt}(t) &=& -{\bf L}_{\rm u}(t)\cdot
{\bf C}_{\rm e}(t)
-{\bf C}_{\rm e}(t)\cdot {\bf L}_{\rm u}^{\top}(t)
\nonumber\\
&& + 2 \Bigl [ {\mathbf \nabla}_{0}{\bf r}_{\rm u}(t)\Bigr ]^{-1}
\cdot {\mathbf \nabla}_{0}{\bf r}(t)\cdot {\bf D}(t)\cdot
\Bigl [ {\mathbf \nabla}_{0}{\bf r}(t)\Bigr ]^{\top}\cdot
\Bigl [ {\mathbf \nabla}_{0}{\bf r}_{\rm u}(t)\Bigr ]^{-\top}.
\end{eqnarray}
Bearing in mind that
\[
\frac{d{\bf C}_{\rm e}^{-1}}{dt}(t)=-{\bf C}_{\rm e}^{-1}(t) \cdot
\frac{d{\bf C}_{\rm e}}{dt}(t)\cdot {\bf C}_{\rm e}^{-1}(t),
\]
we find from Eqs. (16) and (25) that
\begin{eqnarray}
\frac{d{\bf C}_{\rm e}^{-1}}{dt}(t) &=&
{\bf C}_{\rm e}^{-1}(t)\cdot {\bf L}_{\rm u}(t)
+{\bf L}_{\rm u}^{\top}(t)\cdot {\bf C}_{\rm e}^{-1}(t)
\nonumber\\
&-& 2 \Bigl [ {\mathbf \nabla}_{0}{\bf r}_{\rm u}(t)\Bigr ]^{\top}
\cdot {\bf C}^{-1}(t)\cdot
{\mathbf \nabla}_{0}{\bf r}(t)\cdot {\bf D}(t)\cdot
\Bigl [ {\mathbf \nabla}_{0}{\bf r}(t)\Bigr ]^{\top}
\cdot {\bf C}^{-1}(t)\cdot {\mathbf \nabla}_{0}{\bf r}_{\rm u}(t).
\end{eqnarray}
The first principal invariant of the tensor ${\bf C}_{\rm e}(t)$
is given by
\[
I_{1}({\bf C}_{\rm e})={\bf C}_{\rm e}:{\bf I},
\]
where the colon stands for convolution.
Differentiating this equality with respect to time and using Eq. (25),
we obtain
\begin{eqnarray*}
\frac{dI_{1}}{dt}\Bigl ({\bf C}_{\rm e}(t)\Bigr ) &=&
-{\bf L}_{\rm u}(t):{\bf C}_{\rm e}(t)
-{\bf C}_{\rm e}(t):{\bf L}_{\rm u}^{\top}(t)
\nonumber\\
&& + 2 \Bigl [ \Bigl ( {\mathbf \nabla}_{0}{\bf r}(t)\Bigr )^{\top}
\cdot \Bigl ( {\mathbf \nabla}_{0}{\bf r}_{\rm u}(t)\Bigr )^{-\top}
\cdot \Bigl ( {\mathbf \nabla}_{0}{\bf r}_{\rm u}(t)\Bigr )^{-1}
\cdot {\mathbf \nabla}_{0}{\bf r}(t)\Bigr ]:{\bf D}(t).
\end{eqnarray*}
Combining this equality with Eq. (14) and introducing the
rate-of-strain tensor for sliding of junctions
\[
{\bf D}_{\rm u}(t)=\frac{1}{2}\Bigl [ {\bf L}_{\rm u}(t)
+{\bf L}_{\rm u}^{\top}(t)\Bigr ],
\]
we arrive at the formula
\begin{equation}
\frac{dI_{1}}{dt}\Bigl ({\bf C}_{\rm e}(t)\Bigr ) =
2 \Bigl [ \Bigl ( {\mathbf \nabla}_{0}{\bf r}(t)\Bigr )^{\top}
\cdot {\bf C}_{\rm u}^{-1}(t) \cdot
{\mathbf \nabla}_{0}{\bf r}(t)\Bigr ]:{\bf D}(t)
-2 {\bf C}_{\rm e}(t):{\bf D}_{\rm u}(t).
\end{equation}
It follows from Eqs. (19) and (27) that
\begin{equation}
\frac{dI_{1}}{dt}\Bigl ({\bf C}_{\rm e}(t)\Bigr ) =
2 \Bigl [ {\bf B}_{\rm e}(t):{\bf D}(t)
- {\bf C}_{\rm e}(t):{\bf D}_{\rm u}(t)\Bigr ] .
\end{equation}
The Cauchy deformation tensors for transitions from the reference
state to the deformed state and from the reference state to the
unloaded state are assumed to obey the incompressibility
condition, which implies that
\[ I_{3} ({\bf C}_{\rm e})=1. \]
It follows from this equality that
\begin{equation}
I_{2}\Bigl ({\bf C}_{\rm e}(t)\Bigr )
=I_{1}\Bigl ({\bf C}_{\rm e}^{-1}(t)\Bigr )
={\bf C}_{\rm e}^{-1}(t):{\bf I}.
\end{equation}
We differentiate Eq. (29) with respect to time, use Eqs. (13), (14)
and (26), and obtain
\begin{eqnarray*}
\frac{dI_{2}}{dt} \Bigl ({\bf C}_{\rm e}(t)\Bigr )
&=& {\bf C}_{\rm e}^{-1}(t):{\bf L}_{\rm u}(t)
+{\bf L}_{\rm u}^{\top}(t): {\bf C}_{\rm e}^{-1}(t)
-2{\bf D}(t):
\nonumber\\
&& \biggl [ \Bigl ( {\mathbf \nabla}_{0}{\bf r}(t)\Bigr )^{\top}
\cdot {\bf C}^{-1}(t)\cdot {\mathbf \nabla}_{0}{\bf r}_{\rm u}(t)
\cdot \Bigl ( {\mathbf \nabla}_{0}{\bf r}_{\rm u}(t)\Bigr )^{\top}
\cdot {\bf C}^{-1}(t)\cdot {\mathbf \nabla}_{0}{\bf r}(t) \biggr ]
\nonumber\\
&=& 2 {\bf C}_{\rm e}^{-1}(t): {\bf D}_{\rm u}(t)
-2 \biggl [ \Bigl ( {\mathbf \nabla}_{0}{\bf r}(t)\Bigr )^{-1}
\cdot {\bf C}_{\rm u}(t)
\cdot \Bigl ( {\mathbf \nabla}_{0}{\bf r}(t)\Bigr )^{-\top}\biggr ]
: {\bf D}(t).
\end{eqnarray*}
This equality together with Eq. (19) results in the formula
\begin{equation}
\frac{dI_{2}}{dt} \Bigl ({\bf C}_{\rm e}(t)\Bigr )
=-2 \Bigl [ {\bf B}_{\rm e}^{-1}:{\bf D}(t)
-{\bf C}_{\rm e}^{-1}(t): {\bf D}_{\rm u}(t)\Bigr ].
\end{equation}
Equations (28) and (30) determine the derivatives of the principal
invariants of the tensor ${\bf C}_{\rm e}(t)$.

Our goal now is to calculate the derivative of an arbitrary smooth
function, $\phi=\phi(I_{1},I_{2})$, of the principal invariants,
$I_{k}$, of the Cauchy deformation tensor, ${\bf C}_{\rm e}(t)$,
with respect to time.
According to the chain rule for differentiation,
\begin{equation}
\frac{d\phi}{dt}=\phi_{,1}(I_{1},I_{2})\frac{dI_{1}}{dt}
+\phi_{,2}(I_{1},I_{2})\frac{dI_{2}}{dt}
\end{equation}
with
\[
\phi_{,k}=\frac{\partial \phi}{\partial I_{k}}.
\]
Substitution of expressions (28) and (30) into Eq. (31) implies
that
\begin{eqnarray}
\frac{d\phi}{dt}
\Bigl ( I_{1}({\bf C}_{\rm e}(t)),I_{2}({\bf C}_{\rm e}(t))\Bigr )
&=& 2 \Bigl [ \phi_{,1}{\bf B}_{\rm e}(t)
-\phi_{,2}{\bf B}_{\rm e}^{-1}(t)\Bigr ]:{\bf D}(t)
\nonumber\\
&& -2 \Bigl [ \phi_{,1}{\bf C}_{\rm e}(t)
-\phi_{,2}{\bf C}_{\rm e}^{-1}(t)\Bigr ]:{\bf D}_{\rm u}(t).
\end{eqnarray}
Equation (32) generalizes the conventional formula for the
derivative of a function $\phi$ of the principal invariants,
$I_{k}$, of the Cauchy deformation tensor ${\bf C}(t)$,
\begin{equation}
\frac{d\phi}{dt}\Bigl ( I_{1}({\bf C}(t)),I_{2}({\bf C}(t))\Bigr )
= 2 \Bigl [ \phi_{,1}{\bf B}(t)-\phi_{,2}{\bf B}^{-1}(t)\Bigr ]
:{\bf D}(t).
\end{equation}
In what follows, we analyze deformation of meso-domains with various potential
energies, $\omega$, where junctions start to slip with respect to the bulk
material at various times, $\tau$, while they moved together with
the bulk medium before the instant $\tau$.
This means that the tensors ${\bf B}_{\rm e}$ and ${\bf C}_{\rm e}$
become functions of three variables, $t$, $\tau$ and $\omega$,
that satisfy the initial conditions
\begin{equation}
{\bf B}_{\rm e}(t, \tau,\omega)\Bigl |_{t=\tau}={\bf B}(\tau),
\qquad
{\bf C}_{\rm e}(t, \tau,\omega)\Bigl |_{t=\tau}={\bf C}(\tau).
\end{equation}
It follows from Eq. (32) that the derivative of an arbitrary function $\phi$
of the first two principal invariants of the Cauchy deformation tensor
${\bf C}_{\rm e}(t,\tau)$ reads
\begin{eqnarray}
&& \hspace*{-20 mm}\frac{\partial \phi}{\partial t}
\Bigl ( I_{1}({\bf C}_{\rm e}(t,\tau,\omega)),
I_{2}({\bf C}_{\rm e}(t,\tau,\omega))\Bigr )
\nonumber\\
&=& 2 \Bigl [ \phi_{,1}{\bf B}_{\rm e}(t,\tau,\omega)
-\phi_{,2}{\bf B}_{\rm e}^{-1}(t,\tau,\omega)\Bigr ]:{\bf D}(t)
\nonumber\\
&&- 2 \Bigl [ \phi_{,1}{\bf C}_{\rm e}(t,\tau,\omega)
-\phi_{,2}{\bf C}_{\rm e}^{-1}(t,\tau,\omega)\Bigr ]:{\bf D}_{\rm u}(t,\tau,\omega).
\end{eqnarray}
Equations (33) and (35) will be used in the next section to calculate
the derivative of the mechanical energy of a filled elastomer with
respect to time.

\section{Strain energy of an elastomer}

We confine ourselves to loading processes that increase the concentration
of active meso-domains with time.
This implies that any active region can be entirely characterized
by its potential energy, $\omega$, and the instant, $\tau$,
when it became active.
We set $\tau=0$ for meso-domains active in the stress-free state.

The conventional affinity hypothesis is adopted for deformation
of passive meso-domains.
This assumption disregards thermal oscillations of junctions
and postulates that the deformation gradient
for the motion of junctions at the micro-level coincides with
that for the movement of appropriate points of an elastomer
at the macro-level (Treloar, 1975).

A passive region is treated as an isotropic incompressible
medium whose mechanical energy per strand, $w$, depends on the
current temperature, $T$ (the strong effect of temperature reflects
the entropic nature of the strain energy for rubbery polymers), the
measure of damage, $\alpha$, and the first two principal invariants,
$I_{k}$, of the right Cauchy deformation tensor ${\bf C}$,
\begin{equation}
w=w\Bigl (\alpha(t),T(t), I_{1}({\bf C}(t)), I_{2}({\bf C}(t))\Bigr ).
\end{equation}
At an arbitrary instant $t\geq 0$, the strain energy of a strand,
$w_{0}$, belonging to a meso-domain with potential energy $\omega$
which became active at instant $\tau\in [0,t]$ is assumed
to be a function of the current temperature, $T$,
and the first two principal invariants of the right Cauchy
deformation tensor ${\bf C}_{\rm e}$,
\begin{equation}
w_{0}=w_{0}\Bigl (T(t), I_{1}({\bf C}_{\rm e}(t,\tau,\omega)),
I_{2}({\bf C}_{\rm e}(t,\tau,\omega))\Bigr ).
\end{equation}
Equations (36) and (37) are based on the conventional hypothesis that
the excluded-volume effect and other multi-chain effects
are screened for an individual chain by surrounding
macromolecules and they may be accounted for in terms of
the incompressibility condition (Everaers, 1998).

To simplify the analysis, the strain energy of a strand is
assumed to be independent of the potential energy, $\omega$,
of a meso-region to which this strand belongs
(the latter means that the functions $w$ and $w_{0}$ are
treated as average mechanical energies of strands).
The mechanical energy of a strand in an active meso-region, $w_{0}$,
differs, however, from the strain energy, $w$, of a strand in
a passive domain.
The mechanical energy of a passive meso-domain with a potential energy
$\omega$ and a damage measure $\alpha$ reads
\begin{equation}
w(\alpha, T, I_{1},I_{2})=w_{0}(T,I_{1},I_{2})+\alpha
\Delta w(T, I_{1}, I_{2}),
\end{equation}
where $\Delta w$ is the increment of strain energy per strand
in a passive meso-domain (with respect to that in an active region)
caused by inter-chain interactions (van der Waals forces).
Equation (38) implies that the strain energy density of an elastomer
is continuous at the point of transition of a passive
meso-domain into the active state (when the damage parameter, $\alpha$,
vanishes).

Summing the mechanical energies of strands belonging to initial
active meso-regions with various potential energies, $\omega$, to passive
meso-domains with various energies, $\omega$, and damage measures, $\alpha$,
as well as to meso-domains that become active at various instants, $\tau\in [0,t]$,
we find the strain energy density per unit mass of an elastomer
\begin{eqnarray*}
W(t) &=& \int_{0}^{\infty} d\omega \biggl [ \int_{0}^{\infty}
\Xi_{\rm p}(t,\alpha,\omega)
w\Bigl (\alpha, T(t),I_{1}({\bf C}(t)), I_{2}({\bf C}(t))\Bigr )d\alpha
\nonumber\\
&& + X_{\rm a}(0,\omega) w_{0}\Bigl (T(t),I_{1}({\bf C}_{\rm e}(t,0,\omega)),
I_{2}({\bf C}_{\rm e}(t,0,\omega))\Bigr )
\nonumber\\
&& + \int_{0}^{t} \gamma (\tau,\omega) w_{0}
\Bigl (T(t),I_{1}({\bf C}_{\rm e}(t,\tau,\omega)),
I_{2}({\bf C}_{\rm e}(t,\tau,\omega))\Bigr ) d\tau \biggr ].
\end{eqnarray*}
Substitution of Eqs. (36) to (38) into this equality results in
\begin{eqnarray}
W(t) &=& N_{\rm p}(t)w_{0}\Bigl (T(t),I_{1}({\bf C}(t)), I_{2}({\bf C}(t))\Bigr )
+Z(t)\Delta w \Bigl (T(t),I_{1}({\bf C}(t)), I_{2}({\bf C}(t))\Bigr )
\nonumber\\
&& + \int_{0}^{\infty} d\omega \biggl  [ X_{\rm a}(0,\omega)
w_{0}\Bigl (T(t),I_{1}({\bf C}_{\rm e}(t,0,\omega)),
I_{2}({\bf C}_{\rm e}(t,0,\omega))\Bigr )
\nonumber\\
&& + \int_{0}^{t} \gamma (\tau,\omega) w_{0}
\Bigl (T(t),I_{1}({\bf C}_{\rm e}(t,\tau,\omega)),
I_{2}({\bf C}_{\rm e}(t,\tau,\omega))\Bigr ) d\tau \biggr ],
\end{eqnarray}
where the function $N_{\rm p}(t)$ is given by Eqs. (6) and (7) and
\begin{equation}
Z(t)=\int_{0}^{\infty} \alpha d\alpha \int_{0}^{\infty} \Xi_{\rm p}
(t,\alpha,\omega) d\omega.
\end{equation}
Our purpose now is to calculate the derivative of the function $W(t)$ with
respect to time.
It follows from Eqs. (7), (33) to (35) and (39) that
\begin{equation}
\frac{dW}{dt}(t)=J(t)\frac{dT}{dt}(t)+2 {\bf \Lambda}(t):{\bf D}(t)-Y_{1}(t)-Y_{2}(t),
\end{equation}
where
\begin{eqnarray}
J(t)&=& N_{\rm p}(t) \frac{\partial w_{0}}{\partial T}
\Bigl (T(t),I_{1}({\bf C}(t)), I_{2}({\bf C}(t))\Bigr )
+Z(t) \frac{\partial \Delta w}{\partial T}
\Bigl (T(t),I_{1}({\bf C}(t)), I_{2}({\bf C}(t))\Bigr )
\nonumber\\
&& + \int_{0}^{\infty} d\omega \biggl  [
X_{\rm a}(0,\omega) \frac{\partial w_{0}}{\partial T}
\Bigl (T(t),I_{1}({\bf C}_{\rm e}(t,0,\omega)),
I_{2}({\bf C}_{\rm e}(t,0,\omega))\Bigr )
\nonumber\\
&& + \int_{0}^{t} \gamma (\tau,\omega) \frac{\partial w_{0}}{\partial T}
\Bigl (T(t),I_{1}({\bf C}_{\rm e}(t,\tau,\omega)),
I_{2}({\bf C}_{\rm e}(t,\tau,\omega))\Bigr ) d\tau \biggr ],
\nonumber\\
{\bf \Lambda}(t) &=& N_{\rm p}(t)\biggl (
w_{0,1}\Bigl (T(t),I_{1}({\bf C}(t)), I_{2}({\bf C}(t))\Bigr ){\bf B}(t)
\nonumber\\
&& -w_{0,2}\Bigl (T(t),I_{1}({\bf C}(t)), I_{2}({\bf C}(t))\Bigr ){\bf B}^{-1}(t)\biggr )
\nonumber\\
&& + Z(t)\biggl (
\Delta w_{,1} \Bigl (T(t),I_{1}({\bf C}(t)), I_{2}({\bf C}(t))\Bigr ){\bf B}(t)
\nonumber\\
&& - \Delta w_{,2} \Bigl (T(t),I_{1}({\bf C}(t)), I_{2}({\bf C}(t))\Bigr ){\bf B}^{-1}(t)\biggr )
\nonumber\\
&& + \int_{0}^{\infty} d\omega \biggl  [ X_{\rm a}(0,\omega) \biggl (
w_{0,1}\Bigl (T(t),I_{1}({\bf C}_{\rm e}(t,0,\omega)),
I_{2}({\bf C}_{\rm e}(t,0,\omega))\Bigr ) {\bf B}_{\rm e}(t,0,\omega)
\nonumber\\
&&- w_{0,2}\Bigl (T(t),I_{1}({\bf C}_{\rm e}(t,0,\omega)),
I_{2}({\bf C}_{\rm e}(t,0,\omega))\Bigr ) {\bf B}_{\rm e}^{-1}(t,0,\omega)\biggr )
\nonumber\\
&& + \int_{0}^{t} \gamma (\tau,\omega) \biggl (
w_{0,1} \Bigl (T(t),I_{1}({\bf C}_{\rm e}(t,\tau,\omega)),
I_{2}({\bf C}_{\rm e}(t,\tau,\omega))\Bigr ) {\bf B}_{\rm e}(t,\tau,\omega)
\nonumber\\
&& -w_{0,2} \Bigl (T(t),I_{1}({\bf C}_{\rm e}(t,\tau,\omega)),
I_{2}({\bf C}_{\rm e}(t,\tau,\omega))\Bigr ) {\bf B}_{\rm e}^{-1}(t,\tau,\omega)\biggr )
d\tau \biggr ],
\nonumber\\
Y_{1}(t) &=& -\frac{d N_{\rm p}}{d t}(t)
w_{0}\Bigl (T(t),I_{1}({\bf C}(t)), I_{2}({\bf C}(t))\Bigr )
\nonumber\\
&& -\frac{dZ}{dt}(t) \Delta w
\Bigl (T(t),I_{1}({\bf C}(t)), I_{2}({\bf C}(t))\Bigr )
\nonumber\\
&& -\int_{0}^{\infty} \gamma(t,\omega)
w_{0} \Bigl (T(t),I_{1}({\bf C}_{\rm e}(t,t,\omega)),
I_{2}({\bf C}_{\rm e}(t,t,\omega))\Bigr ) d\omega,
\nonumber\\
Y_{2}(t) &=& 2\int_{0}^{\infty} \biggl [ X_{\rm a}(0,\omega){\bf \Lambda}_{\rm u}(t,0,\omega):
{\bf D}_{\rm u}(t,0,\omega)
\nonumber\\
&&+\int_{0}^{t} \gamma(\tau,\omega) {\bf \Lambda}_{\rm u}(t,\tau,\omega):
{\bf D}_{\rm u}(t,\tau,\omega)d\tau \biggr ] d\omega .
\end{eqnarray}
The function ${\bf \Lambda}_{\rm u}$ reads
\begin{eqnarray}
{\bf \Lambda}_{\rm u}(t,\tau,\omega) &=&
w_{0,1}\Bigl (T(t),I_{1}({\bf C}_{\rm e}(t,\tau,\omega)),
I_{2}({\bf C}_{\rm e}(t,\tau,\omega))\Bigr ) {\bf C}_{\rm e}(t,\tau,\omega)
\nonumber\\
&&- w_{0,2}\Bigl (T(t),I_{1}({\bf C}_{\rm e}(t,\tau,\omega)),
I_{2}({\bf C}_{\rm e}(t,\tau,\omega))\Bigr ) {\bf C}_{\rm e}^{-1}(t,\tau,\omega).
\end{eqnarray}
Equations (7), (9), (34) and (42) imply that
\begin{equation}
Y_{1}(t) = -\frac{dZ}{dt}(t) \Delta w \Bigl (T(t),I_{1}({\bf C}(t)),
I_{2}({\bf C}(t))\Bigr ),
\end{equation}
which means that the function $Y_{1}$ is non-negative for an
arbitrary program of loading with an increasing number
of active meso-domains.

\section{Constitutive equations}

We confine ourselves to quasi-static loadings with finite strains,
when mechanically induced changes in temperature, $\Delta T$, are
rather weak.
This implies that the effect of temperature on the specific heat,
$c$, may be disregarded.
The following expression is adopted for the free energy (per unit mass)
of a filled elastomer:
\begin{equation}
\Psi=\Psi_{0}+(c-S_{0})(T-T_{0})-c T \ln\frac{T}{T_{0}}+W,
\end{equation}
where $\Psi_{0}$ and $S_{0}$ are the free energy and the entropy
per unit mass in the stress-free state at the reference
temperature $T_{0}$.
The second and third terms on the right-hand side of Eq. (45)
characterize the free energy of thermal motion of chains.

For an incompressible medium, the Clausius--Duhem inequality reads
(Haupt, 2000)
\begin{equation}
Q=-S\frac{dT}{dt}-\frac{d\Psi}{dt} +\frac{1}{\rho}
\Bigl ( {\mathbf \Sigma}^{\prime}:{\bf D}-\frac{1}{T}{\bf q}\cdot {\bf \nabla} T
\Bigr) \geq 0,
\end{equation}
where $\rho$ is mass density,
${\mathbf \Sigma}^{\prime}$ is the deviatoric component of
the Cauchy stress tensor ${\mathbf \Sigma}$,
${\bf q}$ is the heat flux vector,
$S$ is the entropy,
and $Q$ is the internal dissipation per unit mass.
Substitution of Eqs. (41) and (45) into Eq. (46) results in
\begin{eqnarray}
Q &=&-\Bigl ( S-S_{0}-c\ln\frac{T}{T_{0}}+J\Bigr )\frac{dT}{dt}
+\frac{1}{\rho}\Bigl ( {\mathbf \Sigma}^{\prime}-2\rho {\bf \Lambda}\Bigr
):{\bf D}
\nonumber\\
&& +Y_{1}+Y_{2}-\frac{1}{\rho T}{\bf q}\cdot {\bf \nabla} T \geq 0.
\end{eqnarray}
The function $Y_{1}$ describes the energy dissipation driven by transition
of passive meso-domains into the active state,
$Y_{2}$ characterizes the entropy production induced by slippage
of junctions with respect to the bulk material,
whereas the last term in Eq. (47) is responsible for internal dissipation
caused by thermal conductivity.

Because Eq. (47) is to be satisfied for an arbitrary loading, the
expressions in brackets vanish. This implies the formula for the
entropy per unit mass
\[
S(t)=S_{0}+c\ln\frac{T(t)}{T_{0}}-J(t)
\]
and the constitutive equation
\[
{\mathbf \Sigma}(t) = -P(t){\bf I}+2\rho {\bf \Lambda}(t),
\]
where $P$ is pressure.
Substitution of expression (42) into this equality implies the
stress--strain relation
\begin{eqnarray}
{\mathbf \Sigma}(t) &=& -P(t){\bf I}+2\rho \biggl \{ N_{\rm p}(t) \biggl (
w_{0,1}\Bigl (T(t),I_{1}({\bf C}(t)), I_{2}({\bf C}(t))\Bigr ){\bf B}(t)
\nonumber\\
&& -w_{0,2}\Bigl (T(t),I_{1}({\bf C}(t)), I_{2}({\bf C}(t))\Bigr ){\bf B}^{-1}(t)\biggr )
\nonumber\\
&& + Z(t)\biggl (
\Delta w_{,1} \Bigl (T(t),I_{1}({\bf C}(t)), I_{2}({\bf C}(t))\Bigr ){\bf B}(t)
\nonumber\\
&& - \Delta w_{,2} \Bigl (T(t),I_{1}({\bf C}(t)), I_{2}({\bf C}(t))\Bigr ){\bf B}^{-1}(t)\biggr )
\nonumber\\
&& + \int_{0}^{\infty} \biggl  [  X_{\rm a}(0,\omega) \biggl (
w_{0,1}\Bigl (T(t),I_{1}({\bf C}_{\rm e}(t,0,\omega)),
I_{2}({\bf C}_{\rm e}(t,0,\omega))\Bigr ) {\bf B}_{\rm e}(t,0,\omega)
\nonumber\\
&&- w_{0,2}\Bigl (T(t),I_{1}({\bf C}_{\rm e}(t,0,\omega)),
I_{2}({\bf C}_{\rm e}(t,0,\omega))\Bigr ) {\bf B}_{\rm e}^{-1}(t,0,\omega)\biggr )
\nonumber\\
&& + \int_{0}^{t} \gamma (\tau,\omega) \biggl (
w_{0,1} \Bigl (T(t),I_{1}({\bf C}_{\rm e}(t,\tau,\omega)),
I_{2}({\bf C}_{\rm e}(t,\tau,\omega))\Bigr ) {\bf B}_{\rm e}(t,\tau,\omega)
\nonumber\\
&& -w_{0,2} \Bigl (T(t),I_{1}({\bf C}_{\rm e}(t,\tau,\omega)),
I_{2}({\bf C}_{\rm e}(t,\tau,\omega))\Bigr ) {\bf B}_{\rm e}^{-1}(t,\tau,\omega)
\biggr ) d\tau \biggr ] d\omega \biggr \} .
\end{eqnarray}
We adopt the Fourier law for the heat flux vector ${\bf q}$,
\[
{\bf q}=-\kappa {\bf \nabla}T,
\]
where $\kappa>0$ is thermal diffusivity.
It follows from this equation that the last term on the right-hand side
of Eq. (47) in non-negative.

The incompressibility condition implies that the first invariant of the
rate-of-strain tensor ${\bf D}_{\rm u}$ vanishes, which means that
\[
{\bf \Lambda}_{\rm u}:{\bf D}_{\rm u}
={\bf \Lambda}_{\rm u}^{\prime}:{\bf D}_{\rm u},
\]
where the prime stands for the deviatoric component of a tensor.
By analogy with Eq. (1), we postulate that the rate-of-strain tensor for
sliding of junctions, ${\bf D}_{\rm u}$, is proportional to the tensor
${\bf \Lambda}_{\rm u}^{\prime}$,
\begin{equation}
\rho {\bf \Lambda}_{\rm u}^{\prime}(t,\tau,\omega)
=\eta {\bf D}_{\rm u}(t,\tau,\omega),
\end{equation}
where $\eta>0$ is a viscosity.
Equation (49) provides the flow rule for slippage of junctions
with respect to the bulk material, which ensures that the term
${\bf \Lambda}_{\rm u}:{\bf D}_{\rm u}$ is non-negative.
It follows from this result and Eq. (42) that $Y_{2}\geq 0$
for an arbitrary program of straining.
Because the function $Y_{1}$ is non-negative as well,
governing equations (48) and (49) guarantee that the dissipation
inequality (47) is fulfilled.

It follows from Eqs. (2), (4), (42) and (49) that
\begin{eqnarray}
{\bf D}_{\rm u}(t,\tau, \omega) &=& \frac{\rho \Gamma_{0}}{G} \exp (-\omega)
\biggl [ w_{0,1} \Bigl (T(t),I_{1}({\bf C}_{\rm e}(t,\tau,\omega)),
I_{2}({\bf C}_{\rm e}(t,\tau,\omega))\Bigr ) {\bf C}_{\rm e}^{\prime}(t,\tau,\omega)
\nonumber\\
&& -w_{0,2} \Bigl (T(t),I_{1}({\bf C}_{\rm e}(t,\tau,\omega)),
I_{2}({\bf C}_{\rm e}(t,\tau,\omega))\Bigr )
\Bigl ( {\bf C}_{\rm e}^{-1}(t,\tau,\omega)\Bigr )^{\prime}\biggr ].
\end{eqnarray}
Formulas (48) and (50) entirely determine the evolution of an ensemble of
meso-regions in a filled elastomer.
These equations may be noticeably simplified provided that
the conventional theory of rubber elasticity is applied
to the description of the mechanical energy in active meso-domains.
Setting
\begin{equation}
\rho w_{0}=G(I_{1}-3),
\end{equation}
which corresponds to the strain energy density of a neo--Hookean material,
we find that
\begin{eqnarray}
{\mathbf \Sigma}(t) &=& -P(t){\bf I}+ 2G \biggl \{ N_{\rm p}(t){\bf B}(t)
+\int_{0}^{\infty} \biggl  [  X_{\rm a}(0,\omega){\bf B}_{\rm e}(t,0,\omega)
\nonumber\\
&& + \int_{0}^{t} \gamma (\tau,\omega) {\bf B}_{\rm e}(t,\tau,\omega)d\tau
\biggr ] d\omega \biggr \}
\nonumber\\
&& +2\rho Z(t)\biggl [
\Delta w_{,1} \Bigl (T(t),I_{1}({\bf C}(t)), I_{2}({\bf C}(t))\Bigr ){\bf B}(t)
\nonumber\\
&& - \Delta w_{,2} \Bigl (T(t),I_{1}({\bf C}(t)), I_{2}({\bf C}(t))\Bigr )
{\bf B}^{-1}(t)\biggr ]
\end{eqnarray}
and
\begin{equation}
{\bf D}_{\rm u}(t,\tau, \omega) = \Gamma_{0} \exp (-\omega)
{\bf C}_{\rm e}^{\prime}(t,\tau,\omega).
\end{equation}
Constitutive equations (52) and (53) are applied in the next section
to analyze stresses in a bar under tension.

\section{Uniaxial tension of a specimen}

Points of the bar refer to Cartesian coordinates $\{ X_{i} \}$
in the stress-free state and to Cartesian coordinates $\{ x_{i} \}$
in the deformed state, $(i=1,2,3)$.
Uniaxial tension of an incompressible medium is described by the formulas
\begin{equation}
x_{1}=k(t)X_{1}, \qquad x_{2}=k^{-\frac{1}{2}}(t) X_{2}, \qquad
x_{3}=k^{-\frac{1}{2}}(t) X_{3},
\end{equation}
where $k=k(t)$ is the extension ratio.
We assume that transition from the reference state to the unloaded state
is determined by equations similar to Eq. (54),
\begin{equation}
\xi_{1}=k_{\rm u}(t,\tau,\omega)X_{1}, \qquad
\xi_{2}=k_{\rm u}^{-\frac{1}{2}}(t,\tau,\omega) X_{2}, \qquad
\xi_{3}=k_{\rm u}^{-\frac{1}{2}}(t,\tau,\omega) X_{3},
\end{equation}
where $\{ \xi_{i}\}$ are Cartesian coordinates in the unloaded configuration
and $k_{\rm u}(t,\tau,\omega)$ is a function to be found.
The Cauchy deformation tensors ${\bf B}(t)$ and ${\bf C}(t)$ read
\begin{equation}
{\bf B}(t) ={\bf C}(t) = k^{2}(t){\bf e}_{1}{\bf e}_{1}+k^{-1}(t)
({\bf e}_{2}{\bf e}_{2}+{\bf e}_{3}{\bf e}_{3}),
\end{equation}
where ${\bf e}_{i}$ are base vectors of the frame $\{ X_{i} \}$.
The Cauchy deformation tensors ${\bf B}_{\rm e}(t,\tau,\omega)$
and ${\bf C}_{\rm e}(t,\tau,\omega)$ are given by
\begin{equation}
{\bf B}_{\rm e}(t,\tau,\omega)={\bf C}_{\rm e}(t,\tau,\omega) =
\biggl (\frac{k(t)}{k_{\rm u}(t,\tau,\omega)}\biggr )^{2}
{\bf e}_{1}{\bf e}_{1}+\frac{k_{\rm u}(t,\tau,\omega)}{k(t)}
({\bf e}_{2}{\bf e}_{2}+{\bf e}_{3}{\bf e}_{3}),
\end{equation}
which implies that
\begin{equation}
{\bf C}_{\rm e}^{\prime}(t,\tau,\omega)=\frac{2}{3}\biggl [
\biggl (\frac{k(t)}{k_{\rm u}(t,\tau,\omega)}\biggr )^{2}
-\frac{k_{\rm u}(t,\tau,\omega)}{k(t)}\biggr ]
\Bigl [ {\bf e}_{1}{\bf e}_{1}-\frac{1}{2}({\bf e}_{2}{\bf e}_{2}
+{\bf e}_{3}{\bf e}_{3})\Bigr ].
\end{equation}
It follows from Eq. (55) that
\begin{equation}
{\bf D}_{\rm u}(t,\tau,\omega)=\frac{1}{k_{\rm u}(t,\tau,\omega)}
\frac{\partial k_{\rm u}}{\partial t}(t,\tau,\omega)
\Bigl [ {\bf e}_{1}{\bf e}_{1}-\frac{1}{2}({\bf e}_{2}{\bf e}_{2}
+{\bf e}_{3}{\bf e}_{3})\Bigr ].
\end{equation}
Substitution of expressions (58) and (59) into the flow rule (53) implies
that the function $k_{\rm u}(t,\tau,\omega)$ satisfies the differential
equation
\begin{equation}
\frac{1}{k_{\rm u}(t,\tau,\omega)}
\frac{\partial k_{\rm u}}{\partial t}(t,\tau,\omega)
=\Gamma_{\ast}\exp (-\omega) \biggl [
\biggl (\frac{k(t)}{k_{\rm u}(t,\tau,\omega)}\biggr )^{2}
-\frac{k_{\rm u}(t,\tau,\omega)}{k(t)}\biggr ]
\end{equation}
with the initial condition
\begin{equation}
k_{\rm u}(t,\tau,\omega)\Bigl |_{t=\tau}=1.
\end{equation}
The coefficient $\Gamma_{\ast}$ in Eq. (60) reads
\[ \Gamma_{\ast}=\frac{2}{3}\Gamma_{0}. \]
It follows from Eqs. (52), (56) and (57) that the non-zero components,
$\Sigma_{k}$, of the Cauchy stress tensor
\[ {\mathbf \Sigma}=\Sigma_{1}{\bf e}_{1}{\bf e}_{1}
+\Sigma_{2} ({\bf e}_{2}{\bf e}_{2}+{\bf e}_{3}{\bf e}_{3})
\]
are given by
\begin{eqnarray*}
\Sigma_{1}(t) &=& -P(t)+2\rho Z(t)\Bigl [
\Delta w_{,1}k^{2}(t)- \Delta w_{,2}k^{-2}(t)\Bigr ]
+2Gk^{2}(t)\biggl \{ N_{\rm p}(t)
\nonumber\\
&& +\int_{0}^{\infty} \Bigl [ X_{\rm a}(0,\omega)
k_{\rm u}^{-2}(t,0,\omega)
+\int_{0}^{t} \gamma(\tau,\omega) k_{\rm u}^{-2}(t,\tau,\omega)d\tau
\Bigr ] d\omega\biggr \},
\nonumber\\
\Sigma_{2}(t) &=& -P(t)+2\rho Z(t)\Bigl [
\Delta w_{,1}k^{-1}(t)- \Delta w_{,2}k(t)\Bigr ]
+2Gk^{-1}(t)\biggl \{ N_{\rm p}(t)
\nonumber\\
&&+\int_{0}^{\infty} \Bigl [ X_{\rm a}(0,\omega)
k_{\rm u}(t,0,\omega)+\int_{0}^{t} \gamma(\tau,\omega) k_{\rm u}(t,\tau,\omega)d\tau
\Bigr ] d\omega\biggr \},
\end{eqnarray*}
where the arguments of the function $\Delta w$ are omitted.
Excluding the unknown pressure $P(t)$ from the expressions for
$\Sigma_{k}$ with the help of the boundary condition
\[ \Sigma_{2}(t)=0, \]
we find the longitudinal stress
\begin{eqnarray}
\Sigma_{1}(t) &=& 2 \biggl [ GN_{\rm p}(t)+\rho Z(t)\Bigl (
\Delta w_{,1}+k^{-1}(t)\Delta w_{,2}\Bigr )\biggr ]
\Bigl ( k^{2}(t)-k^{-1}(t)\Bigr )
\nonumber\\
&&+ 2G\int_{0}^{\infty} \biggl [ X_{\rm a}(0,\omega)
\biggl ( \Bigl ( \frac{k(t)}{k_{\rm u}(t,0,\omega)}\Bigr )^{2}
-\frac{k_{\rm u}(t,0,\omega)}{k(t)} \biggr )
\nonumber\\
&& +\int_{0}^{t} \gamma (\tau,\omega)
\biggl ( \Bigl (\frac{k(t)}{k_{\rm u}(t,\tau,\omega)}\Bigr )^{2}
-\frac{k_{\rm u}(t,\tau,\omega)}{k(t)} \biggr ) d\tau \biggr ]
d\omega.
\end{eqnarray}
Equations (60) to (62) determine the stress in a specimen for an arbitrary
program of straining.
To compare results of numerical simulation with experimental data,
we focus on a tensile relaxation test with
\begin{equation}
k(t)=\left \{ \begin{array}{lll}
1, && t<0,\\
\lambda, && t>0,
\end{array}
\right .
\end{equation}
where $\lambda >1$ is a constant.

For the loading program (63), $\Delta w_{,1}$ and $\Delta w_{,2}$
become functions of $\lambda$ only,
\[ \Delta w_{,k}=\Delta w_{,k}(\lambda) \qquad (k=1,2). \]
The concentration of active meso-regions is assumed to be altered
at the instant $t=0$ only, which implies that the function $\gamma(t,\omega)$
vanishes for any $t>0$ and the quantities $N_{\rm p}$ and $Z$ are entirely
determined by the elongation ratio $\lambda$,
\[ N_{\rm p}=N_{\rm p}(\lambda), \qquad Z=Z(\lambda). \]
The number, $X_{\rm a}(0,\omega)$, of active meso-domains
with a potential energy $\omega$ becomes a function of $\lambda$,
\[ X_{\rm a}(0,\omega)=N_{\rm a}(\lambda)p(\lambda,\omega), \]
where $N_{\rm a}(\lambda)$ is the number of active regions at stretching
to the elongation ratio $\lambda$, and $p(\lambda, \omega)$ is the
distribution function for active meso-domains that satisfies the condition
\begin{equation}
\int_{0}^{\infty} p(\lambda,\omega) d\omega=1.
\end{equation}
For the relaxation test (63), the quantity $k_{\rm u}(t,0,\omega)$ also
becomes a function of $\lambda$,
\[ k_{\rm u}(t,0,\omega)=h(t,\lambda,\omega). \]
Using Eqs. (10), (11) and introducing the notation,
\begin{eqnarray}
\psi_{1}(\lambda) &=& 2 \biggl [ G N +\rho Z(\lambda)\Bigl (\Delta w_{,1}(\lambda)
+\lambda^{-1}\Delta w_{,2}(\lambda) \Bigr )\biggr ]
\Bigl ( \lambda^{2}-\lambda^{-1}\Bigr ),
\nonumber\\
\psi_{2}(\lambda) &=& 2G N_{\rm a}(\lambda)\Bigl ( \lambda^{2}-\lambda^{-1}\Bigr ),
\end{eqnarray}
we find from Eqs. (62) and (63) that
\begin{equation}
\Sigma_{1}(t,\lambda)=\psi_{1}(\lambda)-\psi_{2}(\lambda)\int_{0}^{\infty}
\Bigl [1-f(t,\lambda,\omega)\Bigr ]p(\lambda,\omega) d\omega,
\end{equation}
where
\begin{equation}
f(t,\lambda,\omega)=\Bigl ( \lambda^{2}-\lambda^{-1}\Bigr )^{-1}
\biggl [ \Bigl (\frac{\lambda}{h(t,\lambda,\omega)}\Bigr )^{2}-
\frac{h(t,\lambda,\omega)}{\lambda} \biggr ].
\end{equation}
It follows from Eqs. (60), (61) and (63) that the function $h(t,\lambda,\omega)$
obeys the differential equation
\begin{equation}
\frac{\partial h}{\partial t}(t,\lambda,\omega) =\Gamma_{\ast}\exp (-\omega) \biggl [
\biggl (\frac{\lambda}{h(t,\lambda,\omega)}\biggr )^{2}
-\frac{h(t,\lambda,\omega)}{\lambda}\biggr ]h(t,\lambda,\omega),
\qquad
h(0,\lambda,\omega)=1.
\end{equation}
Equations (64) and (66) imply that the dimensionless ratio
\[
R(t,\lambda)=\frac{\Sigma_{1}(t,\lambda)}{\Sigma_{1}(0,\lambda)}
\]
is given by
\begin{equation}
R(t,\lambda)=1-A(\lambda)\int_{0}^{\infty}
\Bigl [1-f(t,\lambda,\omega)\Bigr ]p(\lambda,\omega) d\omega,
\end{equation}
where
\begin{equation}
A(\lambda)=\frac{\psi_{2}(\lambda)}{\psi_{1}(\lambda)}.
\end{equation}
To fit experimental data, we employ the quasi-Gaussian
distribution function
\begin{equation}
p(\lambda,\omega) = p_{0}(\lambda)\exp \biggl [
-\frac{(\omega-\langle\omega(\lambda)\rangle_{0})^{2}}{2\sigma^{2}(\lambda)}\biggr
],
\end{equation}
where $\langle\omega\rangle_{0}$ and $\sigma$ are adjustable parameters
and $p_{0}$ is determined by condition (64).
Substitution of expression (71) into Eq. (69) results in
\[
R(t,\lambda)=1-A(\lambda)p_{0}(\lambda) \int_{0}^{\infty}
\Bigl [1-f(t,\lambda,\omega)\Bigr ]
\biggl [ -\frac{(\omega-\langle\omega(\lambda)\rangle_{0})^{2}}{2\sigma^{2}(\lambda)}
\biggr ] d\omega.
\]
Introducing the new variable, $z=\omega-\omega_{\ast}$,
where $\omega_{\ast}=\ln\Gamma_{\ast}$, we obtain
\[
R(t,\lambda)=1-A(\lambda)p_{0}(\lambda) \int_{-\omega_{\ast}}^{\infty}
\Bigl [1-f(t,\lambda,z)\Bigr ] \biggl [ -\frac{(z-\langle\omega(\lambda)\rangle)^{2}}
{2\sigma^{2}(\lambda)} \biggr ] d z,
\]
where $\langle\omega\rangle=\langle\omega\rangle_{0}-\omega_{\ast}$.
Assuming the relative width of the quasi-Gaussian distribution,
\begin{equation}
\zeta=\frac{\sigma}{\langle\omega\rangle}
\end{equation}
to be small compared to $\langle\omega\rangle$ (this hypothesis will
be discussed later), we neglect the integral over the
interval $[-\omega_{\ast},0]$ and find that
\begin{equation}
R(t,\lambda)=1-A(\lambda)p_{0}(\lambda) \int_{0}^{\infty}
\Bigl [1-f(t,\lambda,z)\Bigr ] \biggl [ -\frac{(z-\langle\omega(\lambda)\rangle)^{2}}
{2\sigma^{2}(\lambda)} \biggr ] d z,
\end{equation}
where the function $f(t,\lambda,z)$ is determined by Eq. (67).
According to Eq. (68), the function $h(t,\lambda,z)$ satisfies
the differential equation
\begin{equation}
\frac{\partial h}{\partial t}(t,\lambda,z) =\exp (-z) \biggl [
\biggl (\frac{\lambda}{h(t,\lambda,z)}\biggr )^{2}
-\frac{h(t,\lambda,z)}{\lambda}\biggr ]h(t,\lambda,z),
\qquad
h(0,\lambda,z)=1.
\end{equation}
Equations (73) and (74) are determined by 3 dimensionless parameters:
\begin{enumerate}
\item
the ratio, $A$, of the relaxing stress to the total stress
at the beginning of the relaxation test;

\item
the average potential energy, $\langle\omega\rangle$, for sliding
of junctions in active in meso-domains;

\item
the standard deviation of the potential energy, $\sigma$.
\end{enumerate}

\section{Experimental procedure}

Three series of uniaxial relaxation tests were performed
at room temperature.
Dumbbell specimens were provided by TARRC laboratories (Hertford, UK)
and were used as received.
The tests were executed on specimens of the compound EDS--16 on the base
of natural rubber reinforced by high abrasion furnace black.
The compound contains 45 phr of carbon black with the
mean diameter of filler particles of about 30 nm
(J\"{a}ger and McQueen, 2001).
A detailed formulation of the compound is presented in Table 1.

Experiments were carried out by using a testing machine
designed at the Institute of Physics (Vienna, Austria) and
equipped with a video-controlled system.
To measure the longitudinal strain, two reflection lines were
drawn in the central part of each specimen before loading
(with the distance 7 mm between them).
Changes in the distance between these lines were measured
by a video-extensometer (with the accuracy of about 1 \%).
The tensile force was measured by using a standard loading cell.
The nominal stress was determined as the ratio of the axial force to
the cross-sectional area of a specimen in the stress-free state, 4
mm $\times$ 1 mm.

To analyze the effects of pre-loading and thermal recovery,
the following testing procedure was chosen:
\begin{enumerate}
\item
Any specimen was loaded with the strain rate
$\dot{\epsilon}_{+}=3.0 \cdot 10^{-2}$ s$^{-1}$
up to a given elongation ratio $\lambda$, which was preserved constant
during the relaxation test ($t_{\rm r}=1$ hour), and unloaded with
the strain rate $\dot{\epsilon}_{-}=-\dot{\epsilon}_{+}$.

\item
The specimen was annealed for $t_{\rm a}=24$ hours at room temperature.

\item
The specimen was loaded with the strain rate $\dot{\epsilon}_{+}$
up to the elongation ratio $\lambda_{\max}=4.0$ and unloaded with
the strain rate $\dot{\epsilon}_{-}$.

\item
The specimen was annealed for $t_{\rm a}$ at ambient temperature.

\item
The specimen was loaded with the strain rate $\dot{\epsilon}_{+}$
up to the same elongation ratio, $\lambda$, as in the first test.
This elongation ratio was preserved constant during the relaxation
test, and, afterwards, the specimen was unloaded with the
strain rate $\dot{\epsilon}_{-}$.

\item
The specimen was recovered in an oven for 24 hours at the constant
temperature $T=100$ $^{\circ}$C and cooled down by air to room
temperature for 4 hours.

\item
The specimen was loaded with the strain rate $\dot{\epsilon}_{+}$
up to the same elongation ratio as in the first test.
This elongation ratio was preserved constant during the relaxation test,
and, afterwards, the specimen was unloaded with the strain rate
$\dot{\epsilon}_{-}$.
\end{enumerate}
The relaxation tests were carried out at four elongation ratios:
$\lambda_{1}=2.0$, $\lambda_{2}=2.5$, $\lambda_{3}=3.0$,
and $\lambda_{4}=3.5$.

The longitudinal stress $\sigma$ was measured as a function of
time $t$ (the initial instant $t=0$ corresponds to the
beginning of the relaxation test).
For any elongation ratio, $\lambda$, the dimensionless ratio
$R(t,\lambda)$ is depicted versus the logarithm of time
($\log=\log_{10}$) in Figures 1 to 3.

These figures evidence that the longitudinal stress is noticeable
decreased during the relaxation time, $t_{\rm r}=1$ h.
For virgin specimens, this decrease equals 19.7 \% at the
elongation ratio $\lambda_{1}$ and it reaches 27.9 \% at $\lambda_{4}$.

After pre-loading to $\lambda_{\max}$ and subsequent unloading,
the time-dependent decrease in stress is weakened.
The portion of the longitudinal stress relaxed during $t_{\rm r}$
is estimated as 14.4 \% at the elongation ratio $\lambda_{1}$
and as 20.0 \% at $\lambda_{4}$.

Thermal recovery of specimens at the elevated temperature $T$
for $t_{\rm a}=24$ h returns the specimens to their initial state.
The decrease in the longitudinal stress during $t_{\rm r}=1$ h
is estimated as 19.7 \% at the smallest elongation ratio, $\lambda_{1}$,
and it grows up to 27.0 \% at straining with $\lambda_{4}$.

These changes in the time-dependent response of the CB filled natural
rubber are associated with alternation of the internal structure
of the compound at the micro-level:
\begin{enumerate}
\item
the concentration of active domains per unit mass grows with an increase
in the elongation ratio,

\item
it is noticeably reduced under pre-loading and subsequent unloading,

\item the concentration of active meso-regions (practically) returns
to its initial value after thermal recovery.
\end{enumerate}
To verify these assertions, we determine adjustable parameters in
Eqs. (73) and (74) by matching experimental data and analyze the effect
of the longitudinal ratio, $\lambda$, on the coefficients $A$,
$\langle\omega\rangle$ and $\sigma$.

\section{Comparison with experimental data}

Given an elongation ratio, $\lambda$, the fitting procedure consists in the following.
We fix intervals $[0,\langle\omega\rangle_{\max}]$ and $[0,\sigma_{\max}]$
where the ``best-fit" parameters $\langle\omega\rangle$ and $\sigma$ are
assumed to be found and divide these intervals into $J$ subintervals by
points $\langle\omega\rangle_{j}=j\Delta_{\omega}$ and $\sigma_{j}=j\Delta_{\sigma}$
with $j=1,\ldots,J$, $\Delta_{\omega}=\langle\omega\rangle_{\max}/J$
and $\Delta_{\sigma}=\sigma_{\max}/J$.
For any pair of parameters $\langle\omega\rangle_{i}$ and $\sigma_{j}$,
we find the constant $p_{0}$ from condition (64),
where the integral is evaluated numerically.
Afterwards, we calculate the integral in Eq. (73) and determine
the pre-factor $A=A(i,j)$ [which ensures the best fit of the experimental
curve $R_{\rm exp}(t,\lambda)$] by the least-squares method.
As a measure of deviations between observations and results of numerical analysis,
we chose the function
\begin{equation}
r(i,j)=\sum_{t_{k}} \Bigl [R_{\rm exp}(t_{k},\lambda)
-R_{\rm num}(t_{k},\lambda)\Bigr ]^{2},
\end{equation}
where the sum is calculated over all experimental points depicted in Figures
1 to 3 and the function $R_{\rm num}(t,\lambda)$ is determined by Eq. (73).
Finally, the ``best-fit" parameters $\langle\omega\rangle$ and $\sigma$ are
determined as those that minimize functional (75) on the set
\[
\Bigl \{ \langle\omega\rangle_{i}, \sigma_{j} \qquad (i,j=1,\ldots, J) \Bigr \}.
\]
To ensure high accuracy in the approximation of observations,
after determining the ``optimal" parameters
$\langle\omega\rangle_{i}$ and $\sigma_{j}$, the above procedure is
repeated for the new intervals
$[\langle\omega\rangle_{i-1},\langle\omega\rangle_{i+1}]$ and
$[\sigma_{j-1}, \sigma_{j+1}]$. In the numerical analysis, we set
$\langle\omega\rangle_{\max}=10.0$, $\sigma_{\max}=10.0$ and
$J=20$. The integrals in Eqs. (64) and (73) are calculated by
using Simpson's method with 100 points and the step $\Delta
z=0.2$. For any $z$, the ordinary differential equation (74) is
solved by the Runge--Kutta method with the time step $\Delta
t=0.05$ s.

To assess the effect of straining of the average potential energy for sliding,
$\langle\omega\rangle$, and the standard deviation of energies, $\sigma$,
we plot these quantities versus the first invariant, $I_{1}$, of the Cauchy
deformation tensor ${\bf C}$.
It follows from Eq. (56) that in tensile relaxation tests,
\[
I_{1}=\lambda^{2}+2\lambda^{-1}.
\]
Experimental data are approximated by the linear functions
\begin{equation}
\langle\omega\rangle=\omega_{0}+\omega_{1}(I_{1}-3),
\qquad
\sigma=\sigma_{0}+\sigma_{1}(I_{1}-3),
\end{equation}
where the coefficients $\omega_{k}$ and $\sigma_{k}$ are
determined by the least-squares technique.
These parameters are listed in Table 2.
Figures 4 and 5 show that phenomenological relations (76)
ensure good accuracy of fitting.

To provide some physical basis for Eq. (76), we refer to Eq. (51)
that implies that for a neo--Hookean medium, the difference $I_{1}-3$
may be treated a measure of the mechanical energy of active meso-regions.
Equations (76) show that the parameters $\langle\omega\rangle$ and $\sigma$
of the distribution function (71) for potential energies of
active meso-domains are proportional to their average strain energy.

Equations (67) and (68) imply that $f(0,\lambda,\omega)=1$
and $f(\infty,\lambda,\omega)=0$.
These equalities, together with Eqs. (64) and (66), yield
\begin{equation}
\Sigma_{1}(0,\lambda)=\psi_{1}(\lambda),\qquad
\Sigma_{1}(\infty,\lambda)=\psi_{1}(\lambda)-\psi_{2}(\lambda).
\end{equation}
We associate $\Sigma_{1}(\infty,\lambda)$ with the equilibrium stress,
$\Sigma_{\rm eq}(\lambda)$,
and $\Delta\Sigma(\lambda)=\Sigma_{1}(0,\lambda)-\Sigma_{1}(\infty,\lambda)$
with the relaxing stress.
It follows from Eqs. (70) and (77) that the ratio
\begin{equation}
a(\lambda)=\frac{\Delta\Sigma(\lambda)}{\Sigma_{\rm eq}(\lambda)}
\end{equation}
reads
\[ a(\lambda)=\frac{A(\lambda)}{1-A(\lambda)}. \]
Given an elongation ratio $\lambda$,
we find the dimensionless ratio $a(\lambda)$ from this equality,
where the value of $A(\lambda)$ is determined by fitting observations,
and plot $a$ versus the first invariant, $I_{1}$, of the Cauchy deformation
tensor.
Figure 6 demonstrates that the experimental data are correctly
approximated by the linear relation
\begin{equation}
a=a_{0}+a_{1}(I_{1}-3),
\end{equation}
where the coefficients $a_{k}$ (found by the least-squares algorithm)
are collected in Table 2.

Figure 4 demonstrates that the average potential energy for sliding
of junctions increases with the elongation ratio $\lambda$.
This may be explained by the fact that only meso-domains
with low potential energies are active in the stress-free state,
whereas regions with higher potential energies become involved
in the sliding process provided that sufficiently large tension
is applied to the specimen.
The rate of growth of the average potential energy for sliding,
$\langle \omega\rangle$, with $I_{1}$ is relatively small
for virgin specimens and it noticeably grows after pre-loading.
After thermal recovery, changes in the function $\langle\omega\rangle(I_{1})$
become similar to those for a virgin sample.
The loading--unloading procedure with $\lambda_{\max}=4.0$
results in a decrease of the average potential energy of the ensemble
of active meso-regions for a stress-free specimen (the parameter
$\omega_{0}$ is reduced by 8 \%).
This observation may be explained by (at least) two reasons:
\begin{enumerate}
\item
damage of the rubbery compound during pre-loading ``turns off" some
active regions in a virgin sample that become passive under unloading;

\item
stretching of a specimen to a high elongation ratio and subsequent
unloading increase the rate of slippage of junctions with respect
to the bulk medium.
According to Eq. (68), this mechanically induced acceleration of
the sliding process is tantamount to a decrease in the potential
energy of active meso-regions.
\end{enumerate}
Figure 5 shows that the standard deviation of potential energies, $\sigma$,
increases with the elongation ratio, $\lambda$.
This growth is rather modest for virgin and recovered samples, but it
becomes stronger after pre-loading: the coefficient $\sigma_{1}$
for the pre-loaded specimens exceeds that for the virgin specimens
by a factor of 6.
After stretching of specimens to $\lambda_{\max}$ and subsequent unloading,
the quasi-Gaussian distribution of potential energies becomes
substantially sharper: the standard deviation of energies for a stress-free
rubbery compound, $\sigma_{0}$, is decreased by 40 \% compared to the virgin
material.
This may be explained by the fact that unloading induces transformation
of active meso-domains with relatively high potential energies (which
characterize a ``tail" of the Gaussian function with large values of
$\omega$) into the passive state.
Thermal recovery at an elevated temperature ``activates" a part of
these regions (the corresponding curve in Figure 5 is located higher
than the curve for samples suffered pre-loading), but no total recovery
is observed: the graph of the function $\sigma(I_{1})$ for the recovered
compound does not reach that for the virgin medium.

To evaluate the dimensionless width of the quasi-Gaussian distribution
function (71), we calculate the ratio $\zeta$ with the help of Eq. (72)
and plot it versus the first invariant, $I_{1}$, of the Cauchy deformation
tensor ${\bf C}$ in Figure 7.
Experimental data are approximated by the linear dependence
\begin{equation}
\zeta=\zeta_{0}+\zeta_{1}(I_{1}-3),
\end{equation}
where the constants $\zeta_{0}$ and $\zeta_{1}$ are determined by the
least-squares algorithm.
Figure 7 demonstrates fair agreement between the observations and
the predictions of the phenomenological equation (80) with the parameters
$\zeta_{0}$ and $\zeta_{1}$ listed in Table 2.

According to Figure 7, the dimensionless ratio $\zeta$
is located within the interval between 0.30 and 0.45 for all specimens,
which implies that the width of the distribution function (71)
is relatively small.
The latter may be thought of as a confirmation of the assumptions
employed in the derivation of Eq. (73).
The quantity $\zeta$ slightly decreases with the elongation ratio $\lambda$
for virgin and recovered specimens, and it weakly increases with $\lambda$
for pre-loaded samples.
The maximal width of the quasi-Gaussian distribution is observed for
the virgin material.
The distribution of potential energies for sliding of junctions
in active meso-regions, $p(\omega)$, becomes sharper after pre-loading,
whereas thermal recovery results in widening of this distribution.

The ratio of the relaxing stress to the equilibrium stress, $a$,
is plotted in Figure 6 as a function of the extension ratio, $\lambda$.
The parameter $a$ is strongly influenced by stretching:
for example, it grows by 45 \% for virgin specimens when $\lambda$
increases from 2.0 to 3.5.
Equations (65), (77) and (78) imply that
\begin{equation}
a(\lambda)=\frac{N_{\rm a}(\lambda)}{N_{\rm p}(\lambda)
+\rho G^{-1}Z(\lambda) [\Delta w_{,1}(\lambda)
+\lambda^{-1}\Delta w_{,2}(\lambda) ]}.
\end{equation}
Because the term in brackets in Eq. (81) [the extra stress caused by
inter-molecular interactions in passive meso-domains] is an increasing
function of $\lambda$, the growth of $a(\lambda)$ is associated
with the following phenomena occurring simultaneously:
\begin{enumerate}
\item
micro-damage of passive meso-regions which results in a decrease of
$Z(\lambda)$,

\item
transformation of passive domains into the active state.
The latter is equivalent to an increase in $N_{\rm a}(\lambda)$
and a corresponding decrease in $N_{\rm p}(\lambda)$ that
is connected with $N_{\rm a}(\lambda)$ by the conservation law (10).
\end{enumerate}
To assess the effect of these two sources for changes in $a(\lambda)$,
constitutive relations are necessary for damage of inter-chain links
in passive meso-regions.
A model for rupture of van der Waals links under loading will be
the subject of a subsequent publication.

Stretching of samples to $\lambda_{\max}$ and subsequent unloading
result in a pronounced decrease in $a$ that is partially recovered
after annealing at an elevated temperature.
The severe decrease in $a$ (the ratio for the relaxing stress to the equilibrium
stress in pre-loaded specimens at relatively small strains, $a_{0}$,
equals about 50 \% of its value for virgin samples) reflects the decay
of the relaxation process evidenced by comparison of curves plotted
in Figures 1 and 2.
This decrease in $a$ is assumed to be closely connected
with the Mullins effect observed in loading--unloading tests on
particle-reinforced elastomers (see, e.g., Bouche, 1960, 1961;
Govindjee and Simo, 1992;
Bergstr\"{o}m and Boyce, 1998;
Miehe and Keck, 2000;
Wu and Liechti, 2000,
to mention a few).

It is worth noting a particular case of the model
where the function $\Xi_{\rm p}$ is independent of time
(or the current strain for a time-dependent program of loading).
This implies that the parameters $N_{\rm a}$, $N_{\rm p}$ and $Z$
remain constants, and $\gamma$ vanishes.
Under these assumptions, the stress--strain relations (52) and (53)
are reduced to the constitutive equations for a rheological
model consisting of a nonlinear spring (which is characterized by
the strain energy $\Delta w$ of inter-chain interactions)
connected in parallel with an infinite
array of the Maxwell elements [neo--Hookean springs linked in series with
nonlinear dashpots whose response is described by Eq. (53)].
A variant of the latter model with a finite number of Maxwell's elements
was widely used in the analysis of the viscoelastic and viscoplastic behavior
of filler rubbers, see, e.g., Govindjee and Simo (1992),
Kaliske and Rothert (1998), Miehe and Keck (2000),
Haupt and Sedlan (2001).
Fitting observations in the relaxation tests reveals that this simplified model
does not capture noticeable changes in the parameters
$\langle\omega\rangle$, $\sigma$ and $a$ under stretching.
Our approach may be treated as an extension of the conventional rheological
model, where the number of Maxwell's elements and their viscosities
are affected by mechanical factors.

\section{Concluding remarks}

Constitutive equations have been derived for the time-dependent behavior
of particle-reinforced elastomers at finite strains.
A filled rubber is modelled as an ensemble of meso-domains where
junctions between chains (chemical and physical cross-links
and entanglements) are not fixed in the bulk medium, but they can slip
with respect to it.
A mean-field approximation is employed to develop stress--strain relations:
a complicated micro-structure of an elastomer reinforced
with aggregates of filler is replaced by equivalent networks
of macromolecules in meso-regions with various potential energies
for sliding of junctions.
The distribution of potential energies of these meso-domains reflects
\begin{enumerate}
\item
impurities and local non-homogeneities in the spatial distribution of
a cross-linker (which is typical of both unfilled and filled elastomers),

\item
density fluctuations driven by segregation of short chains to the
interfaces between filler particles and the host medium,

\item
severe micro-stresses in the neighborhoods of filler clusters in a strained
elastomer caused by the strong difference in the elastic moduli of filler
and rubber.
\end{enumerate}
It is assumed that junctions slip with respect to the bulk material in
active meso-domains, and the rate of sliding is determined by Boltzmann's formula.
Sliding of junctions in passive meso-regions is prevented by
surrounding macromolecules (by means of inter-chain links that
are partially ruptured under loading).
The concentration of links that forbid sliding is described by a measure
of micro-damage, $\alpha$.
Activation of a passive meso-domain occurs when the van der Waals forces between
strands vanish that prevent sliding of junctions.

The mechanical energy of a filled elastomer is determined as the sum
of strain energies of chains in active and passive meso-domains
and the energy of interaction between macromolecules in passive regions.
Constitutive equations for a filled elastomer are derived by using the laws
of thermodynamics (the study is confined to the loading processes in which
the concentration of active meso-regions does not decrease with time).
These equations are applied for the analysis of stresses in an incompressible bar
under uniaxial tension.

To validate the model, three series of tensile relaxation tests were performed
on carbon black filled natural rubber at elongations up to 350 \%.
In the first series, virgin specimens are used,
in the second series, relaxation experiments were carried out on the samples
after mechanical pre-loading,
and in the third series, the time-dependent response was measured
on the same specimens after recovery at an elevated temperature.
Figures 1 to 3 demonstrate fair agreement between the experimental data
and the results of numerical simulation.

To describe quantitatively changes in the internal structure of
a particle-reinforced elastomer induced by pre-loading and thermal
recovery, the distribution of potential energies for sliding of junctions
is approximated by a quasi-Gaussian function with two adjustable parameters.
It is revealed that the average potential energy and the standard
deviation of energies grow with the elongation ratio $\lambda$.
After the loading--unloading procedure, these parameters are noticeably
reduced.
The latter is associated with rearrangement of active meso-domains
driven by rupture of filler clusters.
Thermal recovery implies that the parameters of the quasi-Gaussian
distribution attempt to return to their values for virgin specimens,
but they do not reach their initial values after annealing
for $t_{\rm a}=24$ h at $T=100$ $^{\circ}$C.

\subsection*{Acknowledgement}

We would like to express our gratitude to Dr. K. Fuller (TARRC,
UK) for providing us with rubber specimens.
Stimulating discussions with Profs. N. Aksel, A. Boukamel and S. Reese
are gratefully acknowledged.
\newpage

\subsection*{References}

\hspace*{5.5 mm} Aksel, N., H\"{u}bner, Ch., 1996. The influence
of dewetting in filled elastomers on the changes of their
mechanical properties. Arch. Appl. Mech. 66, 231--241.

Bergstr\"{o}m, J.S., Boyce, M.C., 1998. Constitutive modelling of
the large strain time-dependent behavior of elastomers. J. Mech.
Phys. Solids 46, 931--954.

Bueche, F., 1960. Molecular basis for the Mullins' effect. J.
Appl. Polym. Sci. 4, 107--114.

Bueche, F., 1961. Mullins' effect and rubber--filler interaction.
J. Appl. Polym. Sci. 5, 271--281.

Carlier, V., Sclavons, M., Jonas, A.M., Jerome, R., Legras, R.,
2001. Probing thermoplastic matrix--carbon fiber interphases. 1.
Preferential segregation of low molar mass chains to the
interfaces. Macromolecules 34, 3725--3729.

Clarke, S.M., Elias, F., Terentjev, E.M., 2000. Ageing of natural
rubber under stress. Eur. Phys. J. E 2, 335--341.

Dannenberg, E.M., 1975. The effects of surface chemical
interactions on the properties of filler-reinforced rubbers.
Rubber Chem. Technol. 48, 410--444.

Desmorat, R., Cartournet, S., 2001. Thermodynamics modelling of
internal friction and hysteresis of elastomers. In: Besdo, D.,
Schuster, R.H., Ihlemann, J. (Eds.), Constitutive Models for
Rubber II. Swets and Zeitlinger, Lisse, pp. 37--43.

Doi, M., Edwards, S.F., 1986. The Theory of Polymer Dynamics.
Clarendon Press, Oxford.

Drozdov, A.D., 1996. Finite Elasticity and Viscoelasticity. World
Scientific, Singapore.

Drozdov, A.D., 2001. A tube concept in finite viscoelasticity of
rubbers. In: Besdo, D.; Schuster, R.H.; Ihlemann, J. (Eds.),
Constitutive Models for Rubber II. Swets and Zeitlinger, Lisse,
pp. 93--104.

Drozdov, A.D., Dorfmann, A., 2001. Finite viscoelasticity of
filled rubbers: the effects of pre-loading and thermal recovery.
Cond-mat/0110269.

Everaers, R., 1998. Constrained fluctuation theories of rubber
elasticity: general results and an exactly solvable model. Eur.
Phys. J. B 4, 341--350.

Govindjee, S., Simo, J.C., 1992. Mullins' effect and the strain
amplitude dependence of the storage modulus. Int. J. Solids
Structures 29, 1737--1751.

Green, M.S., Tobolsky, A.V., 1946. A new approach to the theory of
relaxing polymeric media. J. Chem. Phys. 14, 80--92.

Ha, K., Schapery, R.A., 1998. A three-dimensional viscoelastic
constitutive model for particulate composites with growing damage
and its experimental validation. Int. J. Solids Structures 35,
3497--3517.

Hansen, D.E., 2000. A mesoscale strength model for silica-filled
polydimethylsiloxane based on atomic forces obtained from
molecular dynamics simulations. J. Chem. Phys. 113, 7656--7662.

Haupt, P., 2000. Continuum Mechanics and Theory of Materials.
Springer-Verlag, Berlin.

Haupt, P., Sedlan, K., 2001. Viscoplasticity of elastomeric
materials: experimental facts and constitutive modelling. Arch.
Appl. Mech. 71, 89--109.

H\"{a}usler, K., Sayir, M.B., 1995. Nonlinear viscoelastic
response of carbon black reinforced rubber derived from moderately
large deformations in torsion. J. Mech. Phys. Solids 43, 295--318.

Herman, M.F., 2001. A length scale dependent model for stress
relaxation in polymer melts. Macromolecules 34, 4580--4590.

Holzapfel, G., Simo, J., 1996. A new viscoelastic constitutive
model for continuous media at finite thermomechanical changes.
Int. J. Solids Structures 33, 3019--3034.

J\"{a}ger; K.-M., McQueen, D.H., 2001. Fractal agglomerates and
electric conductivity in carbon black polymer composites. Polymer
42, 9575--9581.

Johnson, A.R., Stacer, R.G., 1993. Rubber viscoelasticity using
the physically constrained system's stretches as internal
variables. Rubber Chem. Technol. 66, 567--577.

Johnson, A.R., Quigley, J., Freese, C.E., 1995. A
viscohyperelastic finite element model for rubber. Comput. Methods
Appl. Mech. Engng. 127, 163--180.

Kaliske, M., Rothert, H., 1998. Constitutive approach to
rate-independent properties of filled elastomers. Int. J. Solids
Structures 35, 2057--2071.

Karasek, L., Sumita, M., 1996. Characterization of dispersion
state of filler and polymer--filler interactions in rubber--carbon
black composites. J. Mater. Sci. 31, 281--289.

Leblanc, J.L., Cartault, M., 2001. Advanced torsional dynamic
methods to study the morphology of uncured filled rubber
compounds. J. Appl. Polym. Sci. 80, 2093--2104.

Liang, J.Z., Li, R.K.Y., Tjong, S.C., 1999. Effects of glass bead
content and surface treatment on viscoelasticity of filled
polypropylene/elastomer hybrid composites. Polym. Int. 48,
1068--1072.

Lion, A., 1996. A constitutive model for carbon black filled
rubber: experimental investigations and mathematical
representation. Continuum Mech. Thermodyn. 8, 153--169.

Lion, A., 1997. On the large deformation behavior of reinforced
rubber at different temperatures. J. Mech. Phys. Solids 45,
1805--1834.

Lion, A., 1998. Thixotropic behaviour of rubber under dynamic
loading histories: experiments and theory. J. Mech. Phys. Solids
46, 895--930.

Lulei, F., Miehe, C., 2001. A physically--based constitutive model
for finite viscoelastic deformations in rubbery polymers based on
a directly evaluated micro--macro--transition. In: Besdo, D.,
Schuster, R.H., Ihlemann, J. (Eds.), Constitutive Models for
Rubber II. Swets and Zeitlinger, Lisse, pp. 117--125.

Miehe, C., 1995. Entropic thermo-elasticity at finite strains:
aspects of the formulation and numerical implementation. Comp.
Meth. Appl. Mech. Engng. 120, 243--269.

Miehe, C., Keck, J., 2000. Superimposed finite
elastic-viscoelastic-plastoelastic stress response with damage in
filled rubbery polymers. Experiments, modelling and algorithmic
implementation. J. Mech. Phys. Solids 48, 323--365.

Phan-Thien, N., Tanner, R.L., 1977. A new constitutive equation
derived from network theory. J. Non-Newtonian Fluid Mech. 2,
353--365.

Phan-Thien, N., 1978. A nonlinear network viscoelastic model. J.
Rheol. 22, 259--283.

Reese, S., Govindjee, S., 1998. Theoretical and numerical aspects
in the thermo-viscoelastic material behaviour of rubber-like
polymers. Mech. Time-Dependent Mater. 1, 357--396.

Septanika, E.G., Ernst, L.J., 1998. Application of the network
alteration theory for modeling the time-dependent constitutive
behaviour of rubbers. 1. General theory. Mech. Mater. 30,
253--263. 2. Further evaluation of the general theory and
experimental verification. Mech. Mater. 30, 265--273.

Spathis, G., 1997. Non-linear constitutive equations for
viscoelastic behaviour of elastomers at large deformations. Polym.
Gels Networks 5, 55--68.

Treloar, L.R.G., 1975. The Physics of Rubber Elasticity. Clarendon
Press, Oxford.

Wu, J.-D., Liechti, K.M., 2000. Multiaxial and time dependent
behavior of a filled rubber. Mech. Time-Dependent Mater. 4,
293--331.

Yatsuyanagi, F., Suzuki, N., Ito, M., Kaidou, H., 2001. Effects of
secondary structure of fillers on the mechanical properties of
silica filled rubber systems. Polymer 42, 9523--9529.
\newpage

\begin{center}
{\bf Table 1:} Chemical composition of compound EDS--16 (phr by
weight)

\vspace*{6 mm}

\begin{tabular}{@{} l r @{}}
\hline
Formulation & Content      \cr \hline
Natural rubber         & 100\\
Zinc oxide             & 5  \\
Stearic acid           & 2  \\
Carbon black (N330)    & 45 \\
Process oil            & 4.5\\
Antidegradant(HPPD)    & 3  \\
Antiozonant wax        & 2  \\
Accelerator (CBS)      & 0.6\\
Sulphur                & 2.5\\
\hline
\end{tabular}
\end{center}
\vspace*{30 mm}

\begin{center}
{\bf Table 2:} Adjustable parameters of the model

\vspace*{6 mm}

%
\end{center}
\vspace*{10 mm}

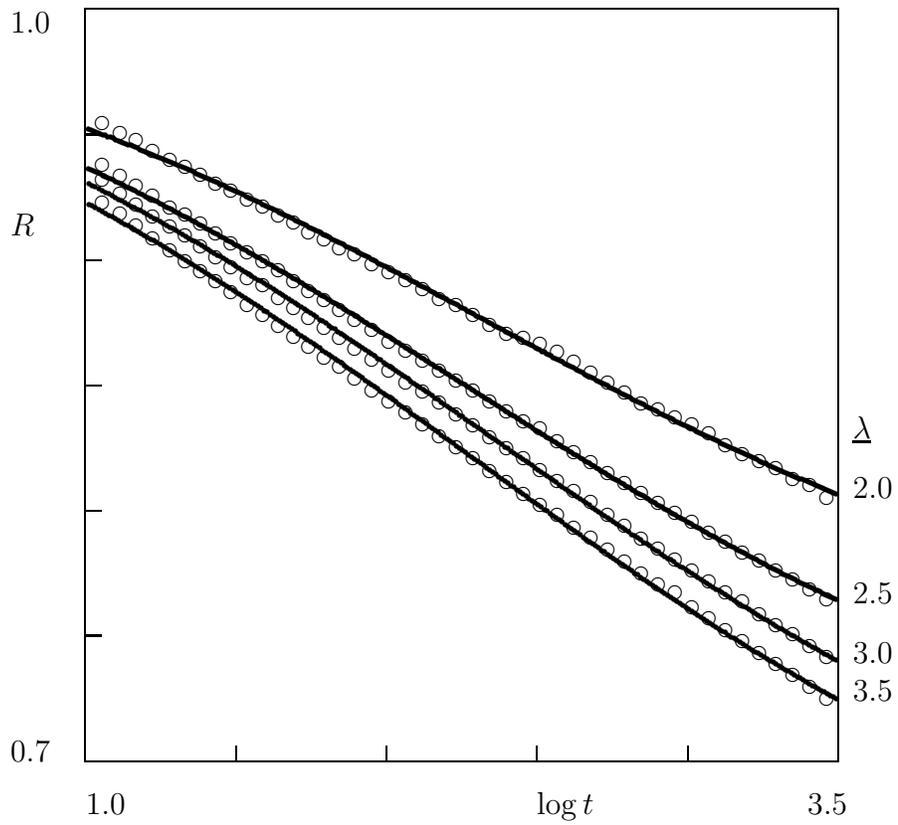
\caption{The dimensionless ratio $R$ versus time $t$ s
in a relaxation test with the elongation ratio
$\lambda$ (virgin specimens).
Circles: experimental data.
Solid lines: results of numerical simulation}
\end{figure}

\begin{figure}[tbh]
\begin{center}
%
\end{center}
\vspace*{10 mm}

\caption{The dimensionless ratio $R$ versus time $t$ s
in a relaxation test with the elongation ratio
$\lambda$ (the specimens after pre-loading).
Circles: experimental data.
Solid lines: results of numerical simulation}
\end{figure}

\begin{figure}[tbh]
\begin{center}
%
\end{center}
\vspace*{10 mm}

\caption{The dimensionless ratio $R$ versus time $t$ s
in a relaxation test with the elongation ratio
$\lambda$ (the specimens after recovery).
Circles: experimental data.
Solid lines: results of numerical simulation}
\end{figure}

\begin{figure}[tbh]
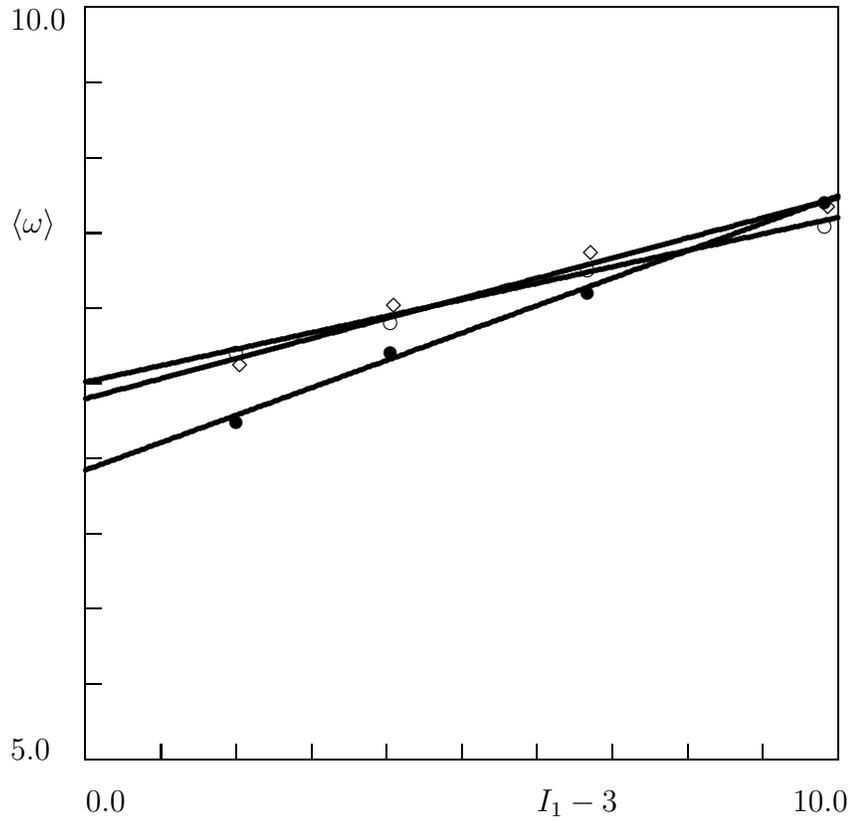

\begin{center}
%
\end{center}
\vspace*{10 mm}

\caption{The average potential energy for sliding of junctions $\langle\omega\rangle$
versus the first invariant $I_{1}$ of the Cauchy deformation tensor.
Symbols: treatment of observations.
Unfilled circles: virgin specimens;
filled circles: the specimens after pre-loading;
diamonds: the specimens after recovery.
Solid lines: approximation of the experimental data by Eq. (76)}
\end{figure}

\begin{figure}[tbh]
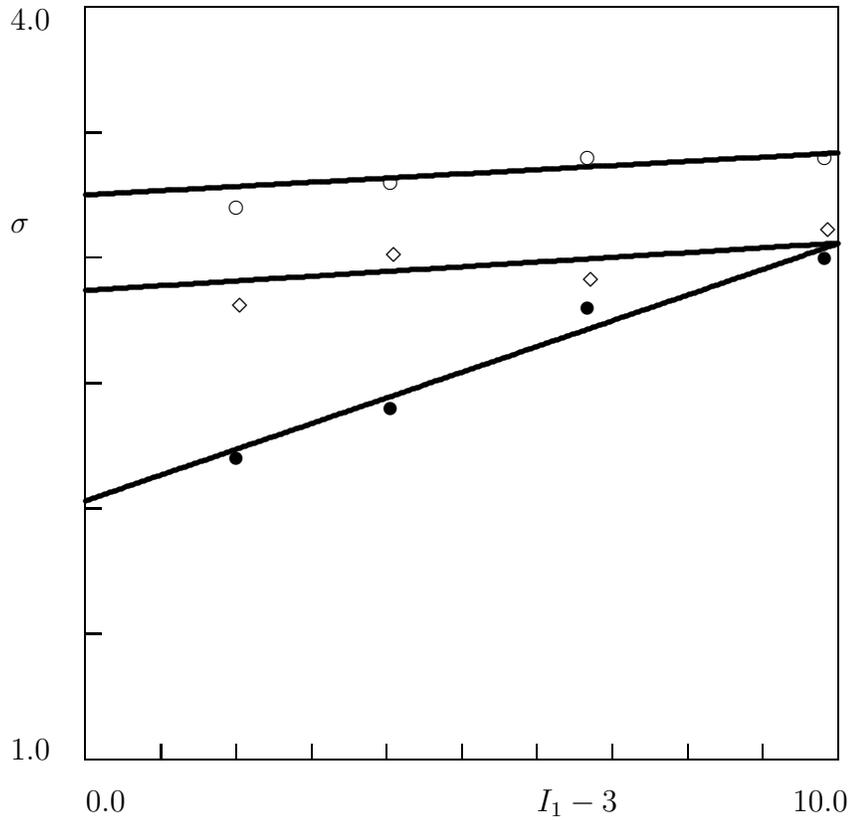

\begin{center}
%
\end{center}
\vspace*{10 mm}

\caption{The standard deviation of potential energies
$\sigma$ versus the first invariant $I_{1}$ of
the Cauchy deformation tensor.
Symbols: treatment of observations.
Unfilled circles: virgin specimens;
filled circles: the specimens after pre-loading;
diamonds: the specimens after recovery.
Solid lines: approximation of the experimental data by Eq. (76)}
\end{figure}

\begin{figure}[tbh]
\begin{center}
%
\end{center}
\vspace*{10 mm}

\caption{The dimensionless parameter $a$ versus the first
invariant $I_{1}$ of the Cauchy deformation tensor.
Symbols: treatment of observations.
Unfilled circles: virgin specimens;
filled circles: the specimens after pre-loading;
diamonds: the specimens after recovery.
Solid lines: approximation of the experimental data by Eq. (79)}
\end{figure}

\begin{figure}[tbh]
\begin{center}
%
\end{center}
\vspace*{10 mm}

\caption{The dimensionless parameter $\zeta$ versus the first
invariant $I_{1}$ of the Cauchy deformation tensor.
Symbols: treatment of observations.
Unfilled circles: virgin specimens;
filled circles: the specimens after pre-loading;
diamonds: the specimens after recovery.
Solid lines: approximation of the experimental data by Eq. (80)}
\end{figure}

\end{document}